\documentclass[preprint,12pt]{elsarticle}

\usepackage{amssymb}
\usepackage{amsmath}
\usepackage{xcolor}
\usepackage{graphicx} 
\usepackage{makeidx}
\usepackage{epsfig}
\usepackage{float}
\usepackage{amscd}
\usepackage{color}
\usepackage{fancyheadings}
\usepackage{fancyvrb}
\usepackage{caption}
\usepackage{hyperref}
\usepackage{subcaption}
\usepackage{cleveref}
\usepackage{verbatim}
\usepackage{diagbox}
\usepackage{multirow}

\graphicspath{{..}}

\newtheorem{example}{Example}

\newcommand{\Real}{\mathbb{R} }

\newcommand{\uu}{{\mathbf u}}
\newcommand{\vv}{{\mathbf v}}
\newcommand{\ww}{{\mathbf w}}
\newcommand{\zz}{{\mathbf z}}
\newcommand{\xx}{{\mathbf x}}
\newcommand{\nn}{{\mathbf n}}
\newcommand{\dz}{{\cal D}(\zz)}

\newcommand{\dv}{{\cal D}(\vv)}

\newcommand{\dvl}{{\cal D}(\vv_\lambda)}
\newcommand{\bfz}{{\mathbf 0}}
\newcommand{\sdiv}{{{\nabla\cdot} \,}}
\newcommand{\gbc}{{\mathbf g}}
\newcommand{\kef}{{\mathcal E}}
\newcommand{\intox}[1]{\int_{\Omega} #1 \, \mathrm{d}\xx}
\newcommand{\bo}{{\partial\Omega}}

\newcommand{\Div}{{\nabla\cdot \,}}

\newcommand{\RE}{{\mathrm{Re}}}

\journal{Nuclear Physics B}

\begin{document}

\begin{frontmatter}



\title{Kinetic energy instability of pipe flow in finite domains}


\author{Chiara Giraudo\corref{XXX}}
\affiliation{organization={University of Oslo},
            country={NO}}
\ead{chiaragiraudo96@gmail.com}
\cortext[XXX]{Corresponding author}

\author{Ingeborg G. Gjerde}
\affiliation{organization={Norwegian Geotechnical Institute},
            country={NO}}
            \ead{ingeborg.gjerde@ngi.no}

\author{Miroslav Kuchta} 
\affiliation{organization={Simula Research Laboratory},
  country={NO}}
  \ead{miroslav@simula.no}

\author{L. Ridgway Scott}
\affiliation{organization={University of Chicago},
  country={USA}}
  \ead{ridg@cs.uchicago.edu }

\begin{abstract}
The instability of pipe flow has been a subject of extensive research, yet a significant gap remains between experimental observations and theoretical predictions. This study revisits the classical problem of kinetic energy instability of pipe flow using contemporary computational methods. We focus on finite domains to address the limitations of previous infinite pipe analyses. We analyze two different cases by imposing homogeneous Dirichlet and periodic boundary conditions. Our investigation reveals that the critical Reynolds number for instability approaches values established by Joseph and Carmi $(1969)$ as the length of the pipe increases. Specifically, when homogeneous Dirichlet boundary conditions are applied, the critical Reynolds number converges monotonically to a value of $\RE_c \leq 81.62$. In contrast, for the periodic case, the critical Reynolds number varies periodically with the length of the pipe, exhibiting local minima at $\RE_c = 81.58$. We further characterize the shape of the perturbations by computing their magnitude and vorticity for several local minima and maxima associated with the periodic problem. Focusing on the perturbations obtained by imposing the homogeneous Dirichlet boundary conditions, we compute their temporal evolution. As expected, the $L^2$-norm of the perturbation decreases from the initial time when solving for the critical Reynolds number, while it initially increases and then eventually decreases for Reynolds numbers exceeding the critical value.
\end{abstract}

\begin{keyword}
{pipe flow \sep instability analysis \sep Reynolds-Orr \sep Taylor-Hood \sep finite element method}



\end{keyword}

\end{frontmatter}


\section{Introduction}
In this study we investigate the critical Reynolds number at which pipe flow becomes unstable. Despite extensive research over many years, a significant gap still exists between experimental and theoretical estimates of this critical value. Linear stability analysis erroneously suggests that pipe flow remains stable for all Reynolds numbers, which is believed to hold true for both axisymmetric and non-axisymmetric disturbances \cite{garg1972linear} \cite[Section 31.2]{drazin2004hydrodynamic}. Experimentally, the onset of instability in pipe flow is identified around Reynolds number $\RE \approx 2000$ \cite{avila2023transition}.

The seminal work of Orr \cite{orr1907stability}, which focused on kinetic energy instability, showed that pipe flow is unstable, but proposed a much lower stability limit at $\RE=180$ for infinitely long pipes. Later, Joseph and Carmi \cite{joseph1969stability} analytically established the existence of an even lower mode at $\RE=82.88$ for perturbations without streamwise variation in infinitely long pipes. In the same paper, they numerically identified a lower stability limit of $\RE=81.49$, associated with a spiral mode. They argued that Orr's mistake was assuming that the most unstable mode is always of a streamwise nature. In their study, the critical Reynolds numbers were obtained by maximizing a functional ratio derived from the Reynolds-Orr energy identity.

Focusing on plane Poiseuille flow, Orszag \cite{orszag1971accurate} showed that it becomes linearly unstable for $\RE >5772$. Recent work by \cite{falsaperla2022energy} supports Orr's assumptions while raising questions about some of Joseph and Carmi's findings \cite{joseph1969stability} with respect to plane Poiseuille flow. In \cite{falsaperla2022energy} the authors examined two types of perturbations in the planar domain $\Real^2 \times [-1,1]$: longitudinal perturbations, independent of $x$ (the streamwise direction), and transverse perturbations, independent of $y$ and with a zero second component in the $y$ direction. Using their notation, Orr assumed that only transverse perturbations were admissible, while Joseph and Carmi also considered longitudinal perturbations when determining the critical Reynolds number. In \cite{falsaperla2022energy}, it was proven that longitudinal perturbations are globally kinematically energy stable for any Reynolds number. This global stability implies that while these perturbations may initially grow, they will eventually decay. Furthermore, the authors conjectured that the maximum should be found on ``a subspace of the space of kinematically
admissible perturbations'', and under this assumption, they established that longitudinal perturbations are kinematically energy stable for all Reynolds numbers. This conjecture was validated for plane Couette and Poiseuille flows in \cite{mulone2024nonlinear}, suggesting that ``the critical nonlinear Reynolds numbers are obtained along two-dimensional perturbations, the transverse perturbations''. {Although} these results hold true for the plane Poiseuille flow, it remains unclear whether and how they can be extended to pipe flow. {In \cite{lrsBIBis}, the Reynolds-Orr instability in planar Couette flow was studied, and while exponential growth was observed for Reynolds numbers up to 100,000, the perturbed flow field ultimately decayed.}

In the context of transient energy growth at subcritical Reynolds number, previous studies \cite{butler1992three, reddy1993energy} for plane Poiseuille flow, along with \cite{schmid1994optimal} for pipe flow, show that longitudinal perturbations characterize the initial conditions that lead to the greatest increase in energy over a specified period of time. This temporary increase in energy is attributed to the non-normal nature of the linear stability operators, where their eigenfunctions are not orthogonal. Furthermore, experimental and simulation studies for Poiseuille flow in cylindrical pipes \cite{avila2023transition, duguet2010slug} support the radial and azimuthal nature of pipe instability. These studies suggest that instability initially appears in localized structures known as puffs and slugs. Puffs occur for $2000 < \RE < 2700$ and are turbulent regions with a defined upstream boundary and a diffuse downstream boundary, maintaining a roughly constant size. In contrast, slugs are observed at $\RE> 3200$ and tend to grow spatially over time. Kerswell \cite{kerswell2005recent} further characterized initial disturbances that can extract energy from Hagen-Poiseuille flow as helical and predominantly streamwise dependent. Recent efforts have aimed to identify alternative solutions to the Navier-Stokes equations to explain the abrupt transition from laminar to turbulent flow in pipes. These solutions, which occur at $\RE>773$ \cite{pringle2007asymmetric}, are thought to facilitate the transition to turbulence. They are characterized by ``wavy streaks with staggered quasistreamwise vortices'' \cite{pringle2007asymmetric} offering a new perspective on pipe flow instability.

Joseph and Carmi \cite{joseph1969stability} employed a Fourier-like transform of the solutions for three-dimensional $(3D)$ Poiseuille flow, based on spiral modes characterized by wave number $\alpha$ and spiral number $N$. They determined the critical Reynolds number by maximizing the functional ratio \cite[p.575]{joseph1969stability}, derived from the Reynolds-Orr equation \cite{ref:SerrinStabilityReOrr}, with respect to $\alpha$ and $N$. Their work suggested that the critical mode could be either periodic or of infinite extend. For this reason, we consider both periodic and non-periodic boundary conditions (BCs) in our study. However, as they remark, the modes obtained from the maximization may not necessarily satisfy the associated Euler-Lagrange equations \cite[p.576]{joseph1969stability}. Our method is based on solving the Euler-Lagrange equations numerically, while weakly enforcing the zero divergence condition. 

With the objective of identifying the most unstable modes within pipe flows across finite domains, this study revisits the Euler-Lagrange equations and develops a finite element method to address the associated eigenvalue problem. Differing from the semi-analytical approach of Joseph and Carmi, which was tailored to pipes, our approach can be applied to arbitrary domains. Notably, our findings show that the critical Reynolds numbers converge toward those previously established by Joseph and Carmi as the length of the pipes increases. Specifically, we identify $\RE_c = 81.58$ as the critical limit for periodic boundary conditions and $\RE_c \leq 81.62$ for homogeneous Dirichlet boundary conditions.  

The paper is organized as follows: In \Cref{sec: Kinetic Energy Instability}, we state the base flow equations
and derive the eigenvalue problem associated with the kinetic energy instability of pipe flow. In \Cref{sec: Numerical Method}, we detail the discretization and numerical methods used to solve the eigenvalue problem and the dynamical problem of the perturbed solution. We discuss our findings in \Cref{sec: results and discussion}. Finally, we summarize our results in \Cref{sec: conclusion}.

\section{Kinetic Energy Instability}
\label{sec: Kinetic Energy Instability}
In this section we report the Navier-Stokes equations that describe the base flow
and derive the eigenvalue problem associated with the kinetic energy stability approach.
Finally we link the eigenvalues and the critical Reynolds number.

\subsection{Base flow equations}
\label{sec:baseflow}
Let $\Omega$ denote a pipe with radius $R$ and length $L$, defined as
\[
    \Omega = \{ \xx=(x,y,z) \in \mathbb{R}^3 : x^2+y^2 < R, \, 0 < z < L\},
\]
with the boundaries denoted as $\Gamma_\text{wall} = \{ \xx \in \partial\Omega : x^2+y^2 = R, 0 \leq z \leq L\}$,
$\Gamma_\text{in} = \left\{ \xx \in \partial\Omega \, : \, z = 0 \right\}$ and
$\Gamma_\text{out}= \left\{ \xx \in \partial\Omega \, : \, z = L \right\}$.
Suppose that the velocity-pressure pair $(\uu,p)$, $\uu:\Omega_T\rightarrow\mathbb{R}^3$, $p:\Omega_T\rightarrow\mathbb{R}$, $\Omega_T = [0, T] \times \Omega$,
solves the dimensionless Navier-Stokes equations in this domain
\begin{equation}\label{eq: base flow eq}
\begin{aligned} 
\frac{\partial{\uu}}{\partial t} - \frac{1}{\RE} \Delta \uu + \,\uu\cdot\nabla \uu 
         + \nabla p &= \bfz \text{ in }\Omega_T,\\
\sdiv\uu &=0 \text{ in } \Omega_T,
\end{aligned}
\end{equation}
together with the initial condition
\begin{equation}
    \uu(0, \xx)=\uu_0(\xx) \quad \forall \xx\in\Omega.
\label{eq: base flow IC}
\end{equation}
We will consider two sets of boundary conditions. In the Dirichlet
case we let
\begin{equation}\label{eq: bcs wall}
  \begin{aligned}
    \uu&=\bfz \quad \text{on } \Gamma_{\text{wall},T} = [0, T] \times \Gamma_\text{wall},\\
    \uu&=\gbc \quad \text{on } \Gamma_{\text{in}, T},\\
    \uu&=\gbc \quad \text{on } \Gamma_{\text{out}, T},\\
  \end{aligned}
\end{equation}
while in case of periodic boundary conditions we prescribe
\begin{equation}\label{eq: bcs periodic} 
\begin{aligned} 
    \uu&=\bfz \text{ on } \Gamma_{\text{wall},T},\\
    \uu(t, \xx_\text{in}) &= \uu(t, \xx_\text{in} + (0,0,L)) \quad \forall (t, \xx_\text{in}) \in \Gamma_{\text{in}, T}.
\end{aligned}
\end{equation}
We note that with the latter boundary conditions we aim to model flow in infinitely long pipes
under the constraint that the solution will be period.

The dimensionless equations are obtained by scaling the spatial coordinates with respect to the radius $R$,
and the velocity with respect to the maximum velocity $U_0$ of the flow. The coefficient $\RE$ is the Reynolds number,
defined for this flow as 
\begin{align}
    \RE = \frac{U_0 R}{\nu},
\end{align}
where $\nu$ is the kinematic viscosity. At low enough Reynolds numbers, pipe flow admits the well-known
steady state solution
\begin{equation}
    \mathbf{u}_{\text{HP}} = (0, 0, u_{\text{HP}}) = \left( 0, 0, U_0 \left( 1- \frac{x^2+y^2}{R^2} \right) \right),
\label{eq: HP flow}
\end{equation}
referred to as Hagen-Poiseuille (HP) flow (e.g. \cite{white2006viscous}). Notably, since the HP flow does not depend
on the $z$-component, it also satisfies the periodic boundary conditions \eqref{eq: bcs periodic}. We next
define the notion of stability of the HP flow.

\subsection{Kinetic energy instability}
\label{sec:kinetic}
To define the concept of kinetic energy instability and derive the corresponding eigenvalue problem, we follow the
methodology outlined in \cite[Sections 2-5]{lrsBIBis}. Let $(\mathbf{u}, p)$ be the laminar solution of the Navier-Stokes equations
\eqref{eq: base flow eq}, \eqref{eq: base flow IC} and \eqref{eq: bcs wall}, with the initial condition given by
the Hagen-Poiseuille flow \eqref{eq: HP flow}. We remark that the following derivation equally applies
also to the period boundary conditions \eqref{eq: bcs periodic}.

Let $(\ww,q)$ be a solution of the Navier-Stokes equations with a different
initial condition than $\uu_0$ (used for $(\mathbf{u}, p)$). That is $(\ww, q)$
satisfies 
\begin{equation}
    \ww_0 = \ww(0, \xx) \not= \uu_0(\xx) \text{ on } \Omega, 
\end{equation}
and 
\begin{equation}
\begin{aligned} \label{eq: perturbed ns system}
\frac{\partial\ww}{\partial t} -\frac{1}{\RE} \Delta \ww + \,\ww\cdot\nabla \ww 
         + \nabla q &= \bfz \text{  in } \Omega_T,\\
\sdiv\ww &=0\text{  in }\Omega_T,\\
\ww &=\bfz \text{  on }\Gamma_{\text{wall},T},\\
\ww&=\gbc \text{  on } \Gamma_{\text{in},T} \cup \Gamma_{\text{out},T}. \\
\end{aligned}
\end{equation}
For their difference $\vv = \uu - \ww$ in $\Omega_T$ with  $\vv_0(\xx) = \vv(0, \xx) \not = 0$ we now consider its {\em kinetic energy}
\begin{equation} \label{eqn:newstabcond} 
\kef(t)=\intox{|\uu(t, \xx)-\ww(t, \xx)|^2}.
\end{equation}
Following \cite[Definition 2.1]{lrsBIBiw}, we say that the flow $\uu$ is {\em kinetic energy unstable} at $t=0$
due to the perturbation $\vv_0$ for Reynolds number $\RE$ if
\begin{equation}
\frac{d\phantom{t}}{dt} \kef(0)>0.
\label{eq: initial energy growth}
\end{equation}
If $\kef'(0)\leq 0$ for all $\vv_0\in V$, then we say that the flow $\uu$ is {\em kinetic energy stable} at $t=0$. Throughout this paper, when we refer to stable or unstable, we mean \textit{kinetic energy stable} or \textit{kinetic energy unstable}, respectively. 

\subsection{The kinetic energy eigenvalue problem}
To derive the eigenvalue problem related to the kinetic energy stability let us first note
that $\vv$ satisfies the equations
\begin{equation}
  \begin{aligned}
    \frac{\partial \vv}{\partial t} -\frac{1}{\RE} \Delta \vv + \big(\uu\cdot\nabla \uu-\ww\cdot\nabla \ww\big) + \nabla  o &= \bfz \text{ in } \Omega_T,\\
\sdiv\vv &= 0 \text{ in } \Omega_T,\\
\vv&=\bfz\;\text{ on }  \Gamma_{\text{wall},T} \\
\vv &= \bfz \text{ on } \Gamma_{\text{in},T} \cup \Gamma_{\text{out},T} \\
\vv(0, \xx) &= \vv_0(\xx) \text{ in }\Omega,
\end{aligned}
\label{eq: perturbation NS} 
\end{equation}
where $o=p-q$.
Before formulating the variational problem for equations \eqref{eq: perturbation NS}, we rewrite
the nonlinear term in the Navier-Stokes equation as follows, using the definition $\vv = \uu -\ww $,
\[
  \uu \cdot \nabla \uu - \ww \cdot \nabla \ww = \uu \cdot \nabla \vv + \vv \cdot \nabla \uu - \vv \cdot \nabla \vv ,
  \]
we introduce the symmetric gradient of $\vv$, 
$\dv = \frac{1}{2} \left( \nabla \vv + \nabla \vv^T \right)$,
the trilinear form
\begin{equation} 
e(\uu,\vv,\ww)=\intox{(\uu\cdot\nabla\vv)\cdot\ww}
\label{eq: trilinear form c}
\end{equation}
and the divergence-free subspace
\[
V_{\text{div}}= \left\{ \boldsymbol{\varphi} \in (H^1(\Omega))^3 \, : \, \nabla \cdot \boldsymbol{\varphi} = 0 \text{ in } \Omega, \, \boldsymbol{\varphi} = \mathbf{0} \text{ on } \partial \Omega \right\}.
\]

By choosing $\vv\in V_{\text{div}}$ as a test function we obtain from \eqref{eq: perturbation NS} that
\begin{equation}  
\frac{1}{2}\frac{\mathrm{d}}{\mathrm{d}t} \intox{|\vv|^2}+ \frac{2}{Re} \intox{|\dv|^2}
+ \big(e(\uu,\vv,\vv) +e(\vv,\uu,\vv)- e(\vv,\vv,\vv)\big)=0.
\label{eq :var form perturbation}
\end{equation}
Exploiting the well-known result from \cite[Section 20.1.2]{lrsBIBih}
\begin{equation}
    e(\uu,\vv,\ww)=- e(\uu,\ww,\vv)+\oint_{\partial\Omega}
                      \uu\cdot\nn\,(\vv\cdot\ww)\,\mathrm{d}\xx
\end{equation}
and noting that $\vv\in V_{\text{div}}$, we find that $e(\uu,\vv,\vv)=0$, 
and $e(\vv,\vv,\vv)=0$ since $\uu=\bfz$ on $\Gamma_\text{wall}$ and $\vv=\bfz$ on $\bo\backslash\Gamma_\text{wall}$.
Moreover, 
\[
e(\vv,\uu,\vv)=\intox{(\vv\cdot\nabla\uu)\cdot\vv}
=\intox{\vv\cdot (\nabla\uu)\vv}
= \intox{\vv\cdot\mathcal{D}(\uu)\vv}.
\]
Thus, \eqref{eq :var form perturbation} can be rewritten as
\begin{equation} 
\frac{1}{2}\frac{\mathrm{d}}{\mathrm{d}t} \intox{|\vv|^2}=-\frac{2}{Re} \intox{|\dv|^2}
- \intox{\vv\cdot\mathcal{D}(\uu)\vv} ,
\label{eq :Reynolds-Orr equation}
\end{equation}
which corresponds to the Reynolds-Orr equation, as noted in
\cite{ref:SerrinStabilityReOrr} and \cite[Section 5.6.1]{schmid2002stability}.

We note that the term $\int_{\Omega} |\dv|^2 \, d\xx$ in \eqref{eq :Reynolds-Orr equation} is always positive.
However, the sign of the second term is not known a priori. In particular, for a given base
flow there may exist an initial perturbation $\vv_0$ for which the right-hand side in
\eqref{eq :Reynolds-Orr equation} is positive. Specifically, from the definition of kinetic energy instability characterized by equation \eqref{eq: initial energy growth}, this means that the flow $\uu$ is unstable at $t=0$ due to the perturbation $\vv$.

As shown in \cite[Section 5.2]{lrsBIBis} the question of existence of such a perturbation to
the HP flow can be related to an eigenvalue problem
\begin{equation}
    b_{\uu}(\vv, \zz) = \lambda a(\vv, \zz) \quad \forall \zz \in V_{\text{div}}.
\label{eq: gen eigenproblem}
\end{equation}
Here the bilinear forms $a, b_\uu$ are defined for any $\vv, \zz \in V_{\text{div}}$ as
\begin{equation}\label{eq:ab_forms}
\begin{aligned}
    a( \vv, \zz) = 2 \intox{\dv:\dz},\quad
    b_{\uu}( \vv, \zz) = \int_\Omega \vv\cdot \mathcal{D}(\uu_0) \zz \,\mathrm{d}\xx.
\end{aligned}
\end{equation}
We recall that $\mathbf{u}_0= \mathbf{u_{\text{HP}}}/U_0$ so that 
\begin{equation}
\mathcal{D}(\uu_0) = 
\frac{1}{2}\left(\nabla \mathbf{u}_0+\nabla \mathbf{u}_{0}^T\right) = \begin{pmatrix}
        0 & 0 & -x \\
        0 & 0 & -y \\
        -x & -y & 0 
    \end{pmatrix}.
\label{eq: M matrix}
\end{equation}
In particular, if there exists a negative $\lambda_c$ such that
\begin{equation}
    \lambda_c = \min_{\bfz\neq\vv\in V_{\text{div}}} \frac{b_{\uu}(\vv, \vv)}{a(\vv, \vv)},
\label{eq: parameter lambda}
\end{equation}
we can equivalently state that the flow $\uu$  is kinematic energy unstable at $t=0$ if 
\begin{equation}
    \RE > -\lambda_c^{-1}.
\label{eq: lambda instability criterion}
\end{equation}
In turn, the critical Reynolds number $\RE_c$ is defined as the Reynolds number at which the flow becomes unstable,
specifically the smallest Reynolds number for which there exists a perturbation $\vv_0 \in V_{\text{div}}$ that
satisfies \eqref{eq: lambda instability criterion} with equality instead of inequality. Thus, with $\lambda$ being
the eigenvalues of \eqref{eq: gen eigenproblem}, the critical Reynolds number can be expressed as
\begin{equation}
    \RE_c = \frac{1}{\max_{\lambda <0} |\lambda|} .
\label{eq: critical Reynolds number}
\end{equation}
The corresponding eigenvector of \eqref{eq: gen eigenproblem} is then the critical perturbation.

\section{Numerical Methods}
\label{sec: Numerical Method}
In this section, we state an alternative form of the eigenvalue problem
\eqref{eq: gen eigenproblem} which allows for practical computations with the
finite element method. We then describe discretization of dynamical problem
employed to study the evolution of the perturbed flow.

\subsection{Computational solution of the eigenproblem}
\label{subsec: comp sol eigenproblem}
Solving the eigenproblem \eqref{eq: gen eigenproblem} for $\vv_{\lambda} \in V_{\text{div}}$ can be
computationally challenging as the basis of $V_{\text{div}}$ is not readily available for
general domain $\Omega$. In the following we shall therefore seek for $\vv_{\lambda}$ in a larger
space $V$ and enforce the divergence-free constraint in a Lagrange multiplier space $Q$. Specifically
we consider the spaces
\begin{equation}
     V= (H_0^1(\Omega))^3, \,Q_0 = L_0^2(\Omega) = \left\{ p \in L^2(\Omega) \, : \, \int_{\Omega} p  \, d\xx = 0\right\}.
\end{equation}
and define the Lagrangian functional
\begin{equation}
    \mathcal{L}(\zz,q) = \frac{1}{2} \left( a(\zz, \zz) -\frac{1}{\lambda} b_{\uu}(\zz, \zz) \right) + c(\zz, q) \, , \quad \zz \in V \, , \, q \in Q_0.
\label{eq: lagrangian functional}   
\end{equation}
where
\begin{equation}
    c(\zz, q) = \int_{\Omega} q \Div \zz \, d\mathbf{x} \qquad \forall \zz \in V, \, \forall q \in Q_0.
\label{eq: d bilinear form}
\end{equation}
Characterizing the saddle point of $\mathcal{L}$, the problem of solving 
the eigenvalue problem \eqref{eq: gen eigenproblem} is equivalent to
finding $\lambda \in \mathbb{R}$, $(\vv_\lambda, p)\in V\times Q_0$ such that
\begin{equation}
\label{eq: variational with pressure}
    \begin{aligned}
        & a(\vv_\lambda, \zz) - \frac{1}{\lambda} \, b_{\uu}(\vv_\lambda, \zz) + c(\zz,p) = 0 \qquad &\forall \zz \in V, \\
        & c(\vv_\lambda, q) = 0  \qquad &\forall q \in Q_0.
	\end{aligned}
\end{equation}
In particular, in the continuous case, due to the choice of spaces, it holds that
$\Div V \subset Q_0$, meaning $\Div \vv_\lambda \in Q_0$. From the second equation in
\eqref{eq: variational with pressure}, this implies that $\Div \vv_\lambda$ is orthogonal
to all functions in $Q_0$, leading to the conclusion that $\Div \vv_\lambda = 0$ almost
everywhere \cite{quarteroni2008numerical}.
We note that in the space $Q_0$, the constraint of zero average over the domain $\Omega$
is necessary to ensure uniqueness in the pressure, since we did not prescribe
any boundary conditions for the pressure. For further details about the well-posedness
of \eqref{eq: variational with pressure} we refer to \cite{boffi2010finite}.

The eigenvalue problem \eqref{eq: variational with pressure} avoids the issue
with $V_{\text{div}}$ space, however, the space $Q_0$ remains impractical for implementation.
Introducing again a larger space $Q=L^2$ and a scalar Lagrange multiplier associated with
a zero mean constraint leads to the final form of the eigenvalue problem:
$\lambda \in \mathbb{R}$, $(\vv_\lambda, p, l)\in V\times Q\times R$ such that
\begin{equation}\label{eq:eigenproblem with lagrange multipliers}
    \begin{aligned}
        & a(\vv_\lambda, \zz)  + c(\zz,p) = \frac{1}{\lambda} \, b_{\uu}(\vv_\lambda, \zz) \qquad &\forall \zz \in V, \\
      & c(\vv_\lambda, q) + d(q, l) = 0  \qquad &\forall q \in Q,\\
      & d(p, m)                    = 0 \qquad &\forall m \in \mathbb{R},
	\end{aligned}
\end{equation}
with $d(\cdot, \cdot)$ denoting the $L^2$-inner product. Introducing operators
\begin{equation}
  \begin{aligned}
& A:\, V \rightarrow V', \quad \big\langle A \vv_\lambda, \zz \big\rangle = 2 \intox{\dvl:\dz}, \\
& B_{\uu}:\, V \rightarrow V', \quad \big\langle B_{\uu} \vv_\lambda, \zz \big\rangle = \int_\Omega (\mathcal{D}(\uu) \vv_\lambda)\cdot \zz \,\mathrm{d}\xx, \\
    & C:\, V \rightarrow Q', \quad \big\langle C \vv_\lambda, p \big\rangle = \int_\Omega p \Div \vv_\lambda \,\mathrm{d}\xx,\\
    & D:\, Q\rightarrow \mathbb{R}, \quad \big\langle D p, l \big\rangle = l\int_\Omega p \,\mathrm{d}\xx.
  \end{aligned}
\end{equation}
we rewrite the eigenvalue problem \eqref{eq: variational with pressure} in the operator form
as: find $\lambda\in\mathbb{R}$, $\mathbf{x}=(\vv_{\lambda}, p, l)\in V\times Q\times\mathbb{R}$ such that
\begin{equation}\label{eq:general GNHEP}
  \mathcal{A}\mathbf{x} = \lambda^{-1}\mathcal{B}\mathbf{x},
\end{equation}
where
\begin{equation}\label{eq:matrices}
\mathcal{A}=
    \begin{bmatrix}
        A & \quad C^T & 0 \\
        C & 0 & D^T \\
        0 & D & 0 
    \end{bmatrix},\quad
\mathcal{B} = 
    \begin{bmatrix}
        B_{\uu} & 0 & 0\\
        0 & 0 & 0 \\
        0 & 0 & 0  
    \end{bmatrix}.
\end{equation}

\subsubsection{Discrete eigenvalue problem}
To discretize \eqref{eq:general GNHEP} we consider triangulation $\mathcal{T}_h$ of
domain $\Omega$ with characteristic mesh size $h>0$ and utilize the lowest order Taylor-Hood
elements, i.e.
\begin{equation}\label{eq:Th}
  \begin{aligned}
    V_h &= \left\{ \vv_h \in (C^0(\Omega))^3: \, \vv_h|_{K} \in ({P}_2(K))^3 \, \forall K \in \mathcal{T}_h \right\} \\
    Q_h &= \left\{ q_h \in C^0(\Omega): \, q_h|_{K} \in P_1(K) \, \forall K \in \mathcal{T}_h \right\}.
  \end{aligned}
\end{equation}
We note that the Taylor-Hood spaces are conforming, $V_h\subset V$ and $Q_h\subset Q$, and
crucially satisfy the inf-sup condition. However, $\Div V_h\not\subseteq Q_h$
and so the divergence-free constraint is not satisfied pointwise.

Denoting $\mathcal{A}_h$, $\mathcal{B}_h$ the matrices due to discretization of
\eqref{eq:matrices} on space \eqref{eq:Th} the discrete version of the eigenproblem
reads: find $\lambda_h \in \Real$ and $\mathbf{x}_h \in V_h \times Q_h \times \Real$ such that 
\begin{equation}
\label{eq: discrete general GNHEP}
    \mathcal{A}_h \mathbf{x}_h = \lambda_h^{-1} \mathcal{B}_h \mathbf{x}_h.
\end{equation}
Here, owing to the inf-sup condition the matrix $\mathcal{A}_h$ is invertible.
We remark that both matrices are symmetric.

In the following we apply Gmsh \cite{geuzaine2009gmsh} to generate tetrahedral
meshes of the computational domain and use FEniCS \cite{logg2012automated} for
the finite element discretization. The eigenvalue problem \eqref{eq: discrete general GNHEP}
is solved with SLEPc \cite{Hernandez:2005:SSF}. Specifically, since the matrices 
$\mathcal{A}_h$ and $\mathcal{B}_h$ in \eqref{eq: discrete general GNHEP} are symmetric but indefinite,
with the left matrix being invertible, the eigenvalue problem is formulated as a (right)
generalized non-Hermitian eigenvalue problem (GNHEP) in SLEPc. To solve it we apply the \textit{shift-and-invert} (SI) spectral transformation, as detailed in the PETSc manual
\cite{dalcinpazklercosimo2011}. Given that the matrix $\mathcal{A}_h$ is non-singular, the scalar
parameter in the SI algorithm is fixed to zero, and we solve the equivalent eigenproblem 
\begin{equation}\label{eq: not-generalized eigenproblem}
    \mathcal{A}_h^{-1}\mathcal{B}_h \mathbf{x}_h= \lambda_h \mathbf{x}_h.
\end{equation}
Here the matrix $\mathcal{A}_h$ is first iteratively inverted using GMRES method, with LU preconditioner.
The discrete eigenproblem \eqref{eq: not-generalized eigenproblem} is then solved using the Krylov-Schur method
\cite{ref:slepcmanual}, which computes the largest eigenvalues in magnitude. Because we employ the spectral method SI,
we require the solver to compute the true residual for the convergence test. During each iteration $ \ \kappa$ of the numerical
method, the relative residual error of the eigenproblem \eqref{eq: discrete general GNHEP},
\begin{equation}\label{eq: eigenproblem residual error}
    r_\kappa = \frac{ \vert \mathcal{A}_h \vv_\kappa - \lambda_{h, \kappa}^{-1} \mathcal{B}_h \vv_\kappa \vert_2}{ \vert \lambda_{h, \kappa}^{-1} \vv_\kappa \vert_2},
\end{equation}
is computed. When this residual is smaller than a predetermined tolerance, the approximating eigenvalue is considered converged. In our simulations, we set this tolerance to be the default value $10^{-8}$. 
Since we are mainly interested in the critical Reynolds number, as defined in \eqref{eq: critical Reynolds number},
we compute only the first few with the largest magnitude. In our experiments, computing 4 extremal eigenvalues was
sufficient to ensure that at least one of them is negative. 

\begin{example}[Verification of eigenvalue solver]\label{ex:solvertest}
To test the proposed solution algorithm we consider a related eigenvalue 
problem with a known solution. In particular, using transformations similar 
to \cite{kobelkov} the eigenvalue problem for the buckling plate $\Omega$
can be shown to be equivalent to solving the problem:
find $(\vv, p, \lambda)\in (H_0^1(\Omega))^2 \times L^2_0\times \mathbb{R}$ such that
\begin{equation}\label{eq:buckling}
\begin{aligned}
-\Delta \vv + \nabla p &= \lambda \vv &\text{ in }\Omega,\\
\nabla\cdot \vv        &= 0 &\text{ in }\Omega,\\
\vv &= 0 &\text{ on }\partial\Omega,
\end{aligned}
\end{equation}
for the smallest eigenvalue $\lambda>0$. Using spectral 
discretization of the original biharmonic problem \cite{bjorstad1999high} 
established that $\lambda\approxeq 52.344691168416544$ for $\Omega=(0, 1)^2$;
a value which will be used as reference in the following.

We observe that \eqref{eq:buckling} differs from \eqref{eq: variational with pressure} 
by using the full gradient operator as opposed to $\mathcal{D}$ in the form $a$. In addition,
for \eqref{eq:buckling} the the bilinear form $b_\uu$ in \eqref{eq:ab_forms} becomes a standard
$L^2$-inner product. After minor modifications which reflect these 
differences we solve \eqref{eq:buckling} by applying the Taylor-Hood spaces \eqref{eq:Th}
and the discrete eigenvalue solver described above.

In \Cref{tab:buckling} we observe that the algorithm converges with 
order 4. This rate agrees with the theoretical estimates \cite[Ch. 13.1]{boffi2010finite} which predict
rate $2s$ for Taylor-Hood elements of order $s$ and $s=2$ for the lowest order 
pair in \eqref{eq:Th}.

\begin{table}
  \begin{minipage}{0.5\textwidth}
        \begin{center}
  \footnotesize{
\begin{tabular}{l|ll}
\hline
$h$ & $\text{dim}W_h$ & $\lvert \lambda - \lambda_h \rvert$\\
\hline
0.125      &    660 & 0.0583287(--)\\
0.0625     &   2468 & 0.00504553(3.53)\\
0.03125    &   9540 & 0.000356594(3.82)\\
0.015625   &  37508 & 2.3388e-05(3.93)\\
0.0078125  & 148740 & 1.49303e-06(3.97)\\
0.00390625 & 592388 & 9.43695e-08(3.98)\\
\hline
\end{tabular}
  }
  \end{center}
\end{minipage}
\begin{minipage}{0.5\textwidth}
  \begin{center}
  \includegraphics[width=0.6\textwidth]{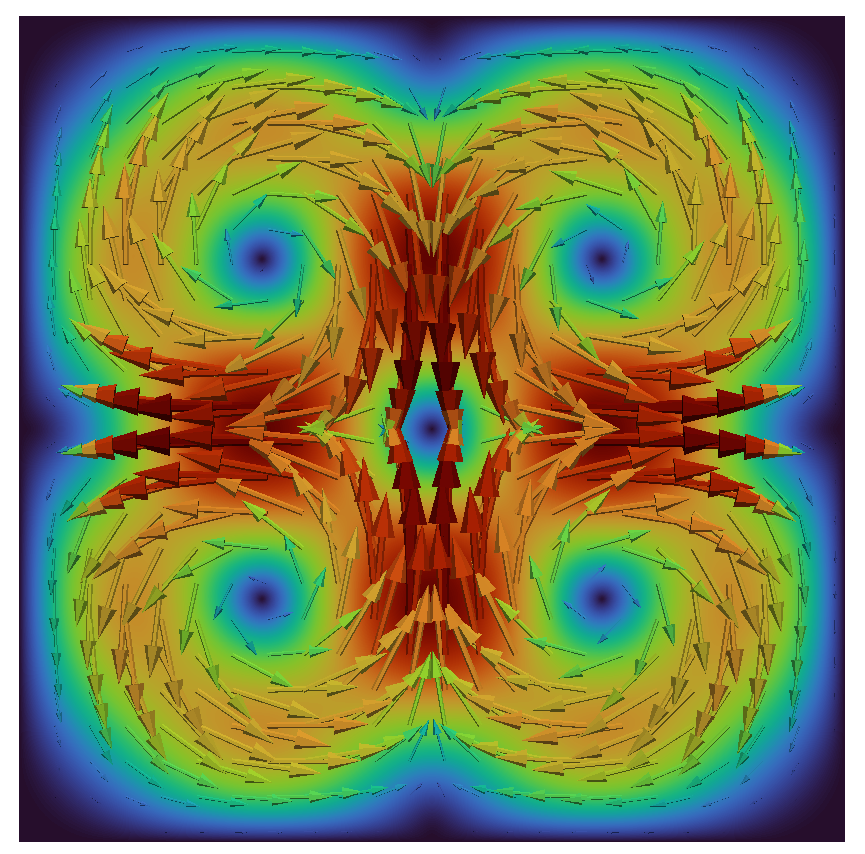}
  \end{center}
\end{minipage}
\caption{
  (Left) Convergence of discrete eigenvalue solver for problem \eqref{eq:buckling} using Taylor-Hood elements
  \eqref{eq:Th} against the result of \cite{bjorstad1999high}. Estimated convergence rate is shown in the brackets.
  (Right) Velocity component of the eigenmode corresponding to the computed eigenvalue.
}
\label{tab:buckling}
\end{table}
\end{example}

\subsection{Time evolution of the perturbation}\label{subsec: comp sol dyn problem}

\begin{figure}[h]
\centering
\begin{subfigure}{.32\textwidth}
  \includegraphics[width=\textwidth]{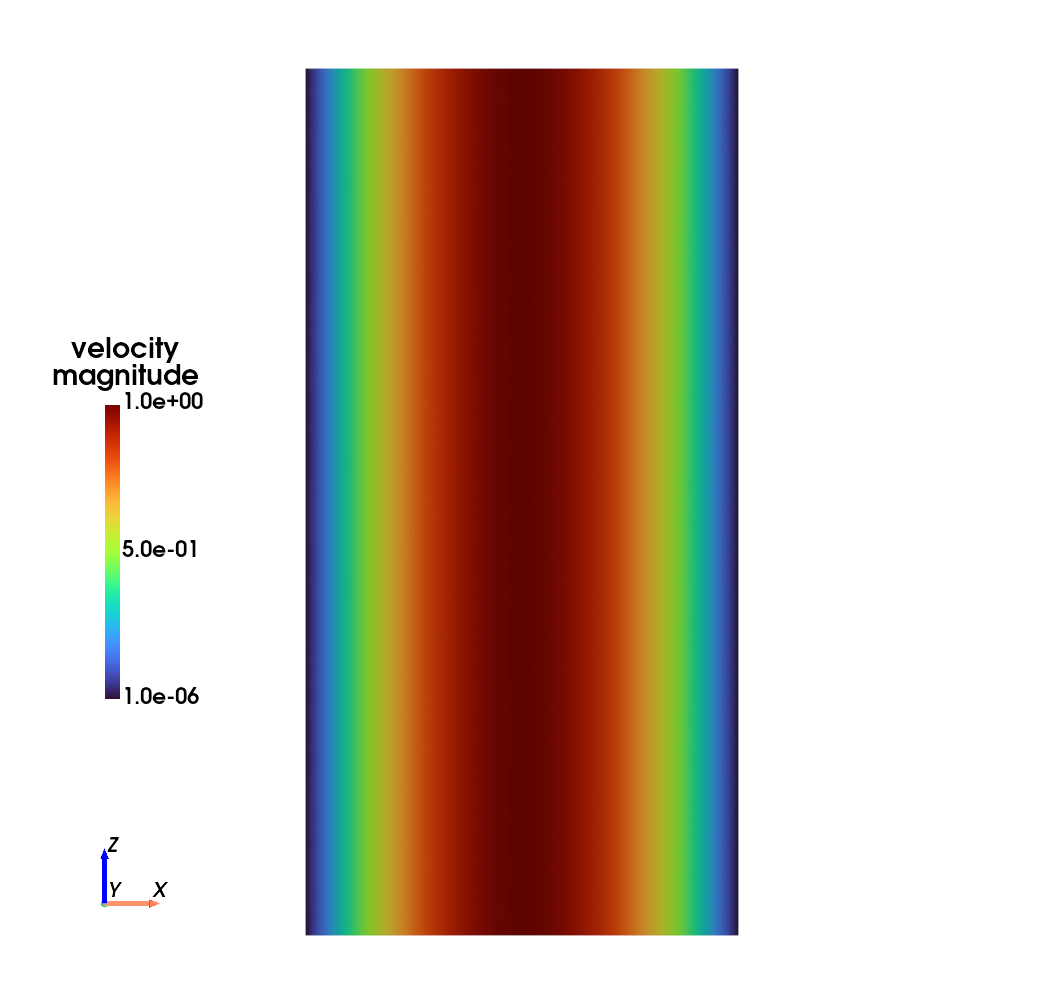}
  \caption{Laminar flow.}
  \label{fig: laminar flow}
\end{subfigure}%
\hfill
\begin{subfigure}{.32\textwidth}
  \includegraphics[width=\textwidth]{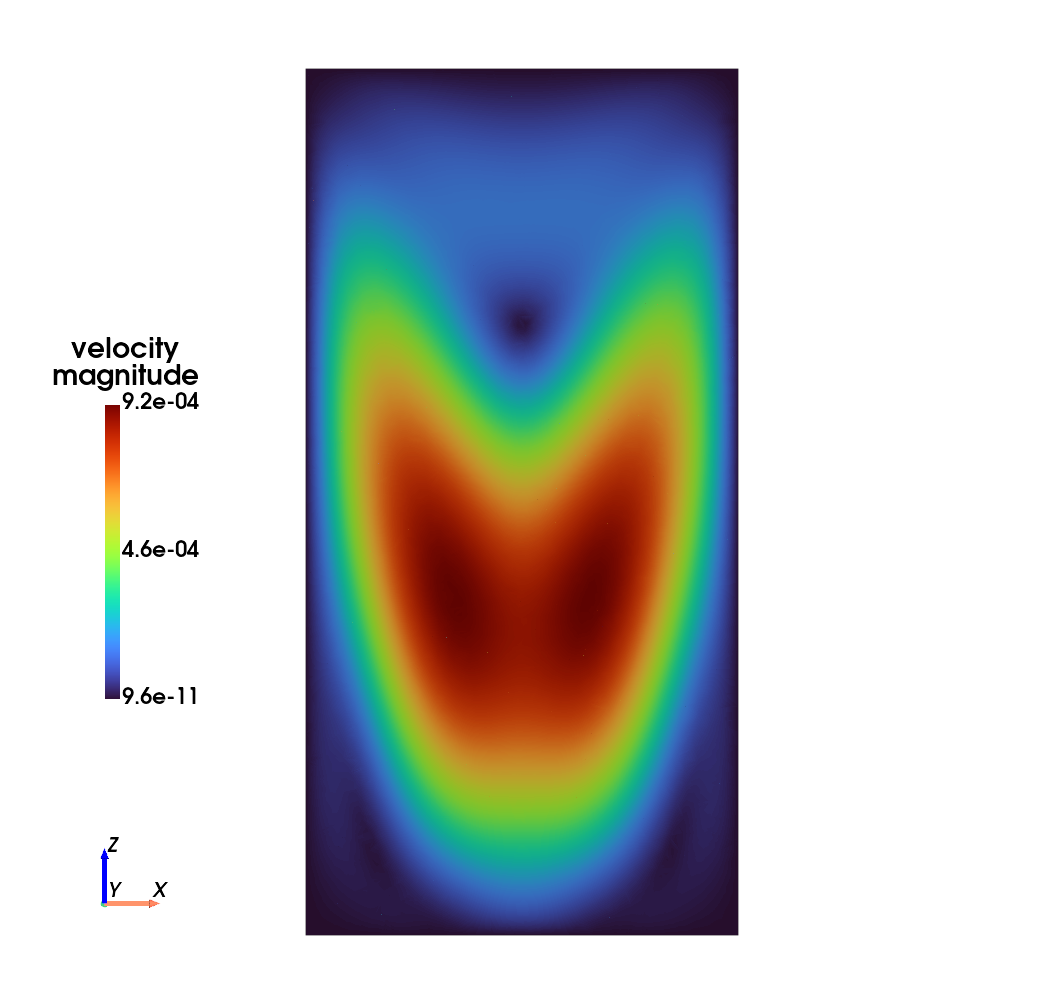}
  \caption{Perturbation with homogeneous Dirichlet BCs.}
  \label{fig: L4 pert homo}
\end{subfigure}
\hfill
\begin{subfigure}{.32\textwidth}
  \includegraphics[width=\textwidth]{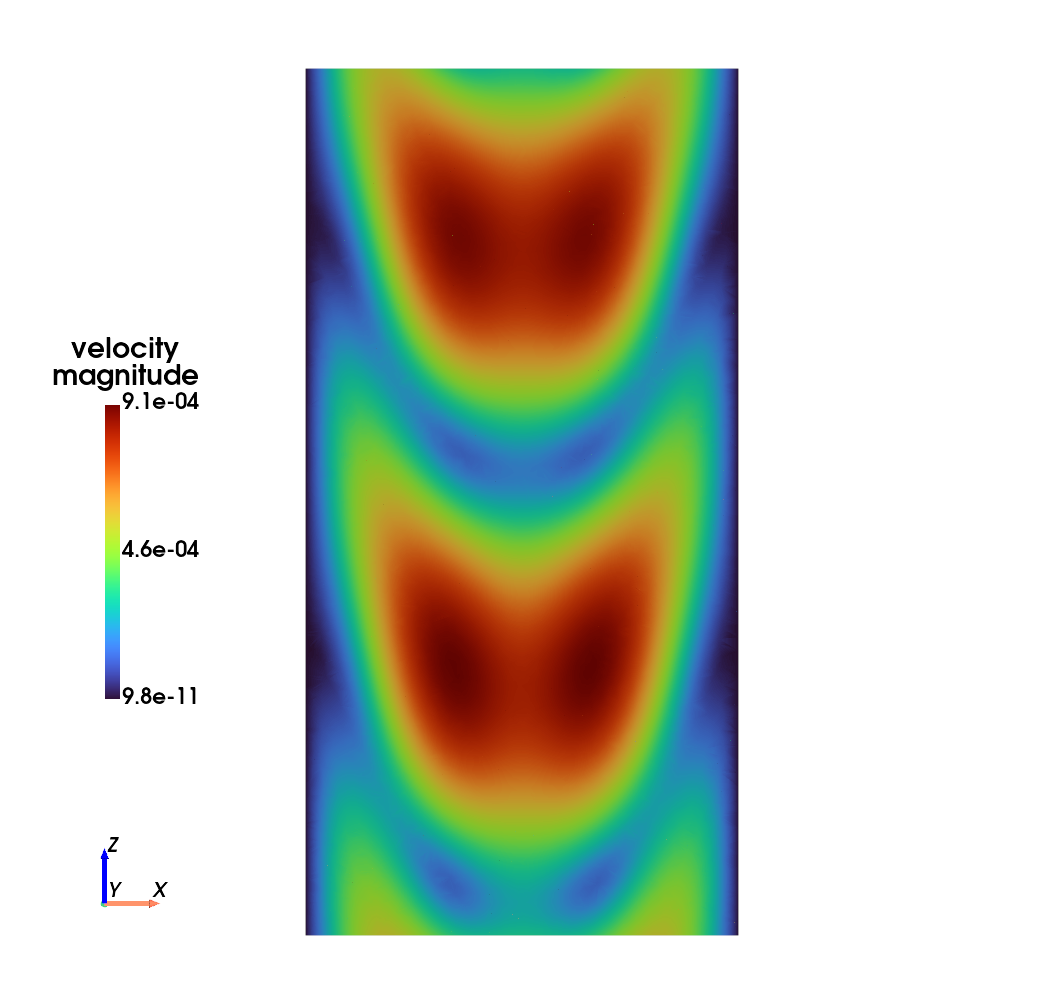}
  \caption{Perturbation with periodic BCs.}
  \label{fig: L4 pert periodic}
\end{subfigure}
\caption{Planar slices of the magnitude of the laminar flow \textbf{(a)} and of the perturbations for homogeneous Dirichlet BCs \textbf{(b)} and periodic BCs \textbf{(c)}, for a pipe length of $L = 4$. The mesh used is the finest, with a size of $h = 0.0625$. The perturbations depicted are those computed by the eigensolver and are associated with the critical Reynolds number.}
\label{fig: ex of laminar and perturbations}
\end{figure}

The temporal evolution of the perturbed flow is described by the system of equations
\eqref{eq: perturbed ns system}, where the boundary conditions are defined as follows
\begin{align}
    \label{eq: perturbed no-slip}
  \ww(t, \xx) &= 0  \quad \text{ on } \Gamma_{\text{wall},T} \\
    \label{eq: perturbed homo Dir BCs}
  \ww(t, \xx) & = \mathbf{u}_{\text{HP}}(\xx) \quad \text{ on } \Gamma_{\text{in}, T} \cup \Gamma_{\text{out}, T}. 
\end{align}
Here $\mathbf{u}_{\text{HP}}$ is the laminar solution, as defined in \eqref{eq: HP flow}. Regarding the initial
conditions, we recall that the perturbed flow $\ww$ is defined as the sum of the laminar solution
$\mathbf{u}_{\text{HP}}$ and a perturbation $\vv$. An example of the perturbed flow is shown in
\Cref{fig: ex of laminar and perturbations} where we present the laminar flow (see \Cref{fig: laminar flow}) alongside the perturbation
under homogeneous Dirichlet boundary conditions (\Cref{fig: L4 pert homo}) and periodic boundary conditions (\Cref{fig: L4 pert periodic}), all obtained for a pipe of length $L=4$. This perturbation is the eigenvector that solves the eigenproblem discussed in
\Cref{sec: Numerical Method}, which is associated with the critical Reynolds number. We recall
that scalar multiple of an eigenvector is also an eigenvector corresponding
to the same eigenvalue. In particular, in our case, due to the normalization by SLEPc,
the eigenvectors computed by the eigensolver typically exhibit a
small magnitude compared to the laminar flow, as seen, for example, in the plot legends of
\Cref{fig: ex of laminar and perturbations}. To facilitate easier comparison
between the eigenmodes we scale the perturbations as
\begin{equation}
    \vv_{\text{scaled}} = \frac{1}{4} \frac{\max_{\xx \in \Omega} \Vert \uu_{\text{HP}}(\xx) \Vert_2 }{\max_{\xx \in \Omega} \Vert \vv_\lambda(\xx) \Vert_2} \vv_\lambda.
\label{eq: scaled perturbation}
\end{equation}
Here, the maximum is taken with respect to the velocity magnitude throughout the domain. The chosen scaling
factor ensures that the perturbation and laminar flow have comparable maximum magnitudes at the initial time,
while still allowing the laminar flow to dominate. Furthermore, consistently applying the same scaling factor
across all simulations enables comparisons of the perturbed flow behavior over time for different scenarios.
Finally, the initial condition is defined as
\begin{equation} 
\label{eq: perturbed scaled ICs}
    \ww(t=0, \xx) = (\mathbf{u}_{\text{HP}} + \vv_{\text{scaled}})(\xx) \quad \text{ in } \Omega, 
\end{equation}
where $\vv_{\text{scaled}}$ is the scaled perturbation defined in \eqref{eq: scaled perturbation} and corresponds
to the most unstable eigenmode.
\subsubsection{Discretization}\label{sec:dyndiscrete}
Numerical solution of the system of equations defined by \eqref{eq: perturbed ns system},
\eqref{eq: perturbed scaled ICs} requires both temporal and spatial discretizations.
To discretize the system in time we consider a uniform partition of 
the interval $[0,T]$ and denote by $\Delta t$ the time step size.
To achieve second-order convergence in time, we apply the Crank-Nicholson (CN) method 
which solves the system of equations at time $t_{n+1/2}$. The time derivative of the velocity at $t_{n+1/2}$ is approximated using
a second-order central scheme, while the velocity at this time is expressed as the average of the known velocity at
time $t_{n}$ and the unknown velocity at time $t_{n+1}$. Following \cite[Chap. 22]{logg2012automated} the convective term
$(\ww_{n+1/2} \cdot \nabla) \ww_{n+1/2}$ can be
approximated either implicitly or semi-implicit. In the implicit case, both velocities in the term are evaluated based on the
approximation of $\ww_{n+1/2} = \ww(t_{n+1/2}, \xx)$, as discussed previously. This leads to the classical Crank-Nicholson
formulation. In the semi-implicit case, the convective term is linearized using Adams-Bashforth projection, where the
convecting velocity is explicitly expressed as a linear combination of the known velocities $\ww_n$ and $\ww_{n-1}$.
This represents the Crank-Nicholson Adams-Bashforth formulation. Both formulations yield a second-order discretization
in time and solve for the velocity at time $t_{n+1}$ and pressure at time $t_{n+1/2}$. The pressure at the subsequent time step
$n+1$ can be derived through post-processing as $q_{n+1} = 2q_{n+1/2} -q_n$. While the initial velocity is known, the
initial pressure is unknown and it set to zero. To improve the pressure approximation, we perform one step of the
Backward Euler method when using the implicit discretization, and two steps of the Backward Euler method when using the semi-implicit discretization. 

Following temporal discretization of \eqref{eq: perturbed ns system} and
\eqref{eq: perturbed scaled ICs} we consider a monolithic scheme where the velocity and pressure
are solved for simultaneously. We use Taylor-Hood elements \eqref{eq:Th} for spatial discretization
such that the scheme is second-order accurate in space. Then, at each discrete time step, the nonlinear
problem is solved computationally through Newton's method with line research, utilizing the SNES library
of PETSc \cite{osti_2565610}. The linear solver employed within Newton's method is set to MUMPS \cite{MUMPS:1},
allowing the code to be run also in parallel.
 
\section{Results and Discussion}
\label{sec: results and discussion}
In this section we examine the relationship between critical Reynolds numbers and
both numerical and physical parameters. The numerical parameter considered is the mesh size,
while the physical parameters include the boundary conditions and the pipe length. We focus our discussion on simulations that have reached convergence and we compare our results with those of Joseph and Carmi \cite{joseph1969stability}. Finally,
we investigate the evolution of the perturbed flow over time for Reynolds numbers equal to or exceeding the critical value.

\subsection{Critical Reynolds number as a function of mesh size and pipe length}
To assess the impact of mesh size on the convergence of the eigensolver, we solved the
discrete eigenproblem for four different mesh sizes, namely $h=0.5, \, 0.25, \, 0.125 \text{ and } 0.0625$.
We evaluated these mesh sizes across various pipe lengths, focusing on the eigenvalues that correspond to
the critical Reynolds numbers $\RE_c$ for each length, as defined by equation \eqref{eq: critical Reynolds number}.
Consequently, we report critical Reynolds numbers rather than eigenvalues. 

In \Cref{tab: Re_c vs L vs mesh_size}, we show the critical Reynolds numbers for different mesh sizes and pipe lengths.
We recall that the eigenvalue/critical Reynolds number is considered converged when the relative residual error,
as defined in \eqref{eq: eigenproblem residual error}, falls below a specified tolerance
($10^{-8}$) and the number of iterations in the eigensolver's Krylov-Schur algorithm does not exceed 100. By
this definition all critical Reynolds numbers in \Cref{tab: Re_c vs L vs mesh_size} are converged. We observe
that the eigenvalues for all lengths are converging with mesh refinement for both choices of boundary conditions.
To assess the error magnitude we consider
in \Cref{tab: relative error vs L} the relative errors $\lvert \RE_{c, 2h} - \RE_{c, h}\rvert/\RE_{c, h}$ for select
pipe lengths $L= 2, 4, 16, 32$. These lengths are provided as examples since
all pipe lengths exhibit similar relative errors, sharing the same order of magnitude.

\begin{table}[!ht]
  \centering
    \scriptsize
\begin{tabular}{|l|llll|llll|}
\hline
& \multicolumn{4}{c|}{Homogeneous Dirichlet BCs} & \multicolumn{4}{c|}{Periodic BCs}   \\ 
\hline
\multicolumn{1}{|l|}{\backslashbox{$L$}{$h$}} & 0.5   & 0.25  & 0.125  & 0.0625 & 0.5   & 0.25  & 0.125  & 0.0625\\
\hline
{2}   & 102.34 & 100.51 & 98.94 & 98.68  & 83.62 & 83.74 & 83.12 & 82.93 \\
{4}   & 85.48 & 83.98 & 83.18 & 82.97    & 83.73 & 83.75 & 82.84 & 82.63 \\
{8}   & 83.29 & 83.08 & 82.23 & 82.02    & 83.23 & 82.82 & 82.04 & 81.84 \\
{12}  & 83.29 & 82.87 & 82.03 & 81.82    & 83.12 & 82.65 & 81.78 & 81.58 \\
{16}  & 82.99 & 82.80 & 81.91 & 81.71    & 82.99 & 82.75 & 81.82 & 81.61 \\
{20}  & 80.93 & 82.76 & 81.87 & 81.66    & 80.92 & 82.69 & 81.85 & 81.64 \\
{24}  & 82.55 & 82.72 & 81.85 & 81.64    & 82.81 & 82.66 & 81.79 & 81.58 \\
{28}  & 81.87 & 82.68 & 81.83 & 81.62    & 81.87 & 82.68 & 81.79 & 81.58 \\
{32}  & 82.98 & 82.70 & 81.82 & 81.62    & 82.97 & 82.65 & 81.82 & 81.61 \\
\hline
\end{tabular}
\caption{Critical Reynolds number with respect to pipe length $(L)$ and mesh size $(h)$.}
\label{tab: Re_c vs L vs mesh_size}
\end{table}

\begin{table}[!ht]
  \centering
    \scriptsize  
\begin{tabular}{|l|lll|lll|}
\hline
& \multicolumn{3}{c|}{Homogeneous Dirichlet BC} & \multicolumn{3}{c|}{Periodic BCs}  \\ 
\hline
\backslashbox{$L$}{$h$} & 0.25  & 0.125  & 0.0625 & 0.25  & 0.125  & 0.0625\\
\hline
{2} & $1.82 \times 10^{-2}$ & $1.58 \times 10^{-2}$ & $2.63 \times 10^{-3}$ 
& $ 1.43 \times 10^{-3}$ & $ 7.45\times 10^{-3}$ & $ 2.29\times 10^{-3}$ \\
{4} & $ 1.78 \times 10^{-2}$ & $ 9.61 \times 10^{-3}$ & $ 2.53 \times 10^{-3}$ 
& $ 2.38\times 10^{-4}$ & $ 1.09\times 10^{-2}$ & $ 2.54 \times 10^{-3}$ \\
{16}  & $ 2.29 \times 10^{-3}$ & $ 1.08 \times 10^{-2}$ & $ 2.44 \times 10^{-3}$ 
& $ 2.90 \times 10^{-3}$ & $ 1.13 \times 10^{-2}$ & $ 2.57 \times 10^{-3}$
\\ 
{32}  & $ 3.38 \times 10^{-3}$ & $ 1.07 \times 10^{-2}$ & $ 2.45 \times 10^{-3}$
& $ 3.87 \times 10^{-3}$ & $ 1.01 \times 10^{-2}$ & $ 2.57 \times 10^{-3}$\\
\hline
\end{tabular}
\caption{Relative error of the select critical Reynolds numbers from \Cref{tab: Re_c vs L vs mesh_size}
  for a fixed pipe length $(L)$; $\lvert \RE_{c, 2h}-\RE_{c, h}\rvert/\lvert \RE_{c, h}\rvert$.
}
\label{tab: relative error vs L}
\end{table}

In \Cref{tab: unknowns vs L vs mesh_size,tab: CPU time vs L vs mesh_size} we report respectively
the number of unknowns in the discrete eigenproblem and the CPU time for obtaining the critical
Reynolds number. Simulations were run in serial on a workstation with AMD EPYC CPU. The reported CPU time represents the total time for assembling the system and
solving the eigenvalue problem. Consistent with previous observations, CPU time increases with mesh
size for fixed pipe lengths, and it also increases with pipe lengths for fixed mesh sizes.
Compared to the number of unknowns, CPU time exhibits greater variability across different cases
(varying pipe lengths or mesh sizes). Nevertheless, these variations generally remain within an order of
magnitude between $10^{-1}$ and $10$. Notably, in the case of pipe length $L=16$ with mesh size $h=0.25$
and periodic boundary conditions, the CPU time is significantly higher than expected. This occurred because
within $100$ iterations only $3$ eigenvalues converged; we require $4$ converged eigenvalues for successful
termination. However, in this case the eigenvalue of interest was actually found after just $17$ iterations.
Among all the cases reported in \Cref{tab: Re_c vs L vs mesh_size}, there are only two instances in which
the CPU time was significantly high due to the convergence of fewer than $4$ eigenvalues within the first $100$ iterations.

\begin{table}[!ht]
  \centering
  \scriptsize
\begin{tabular}{|l|llll|llll|}
\hline
& \multicolumn{4}{c|}{Homogeneous Dirichlet BCs} & \multicolumn{4}{c|}{Periodic BCs}   \\ 
\hline
\multicolumn{1}{|l|}{\backslashbox{$L$}{$h$}} & 0.5   & 0.25  & 0.125  & 0.0625 & 0.5   & 0.25  & 0.125  & 0.0625\\
\hline
{2}   & 5404 & 11797 & 73736 & 521827 
& 4820 & 11013 & 70197 & 508816 \\
{4}   & 4174 & 21434 & 142623 & 1023735
& 3871 & 20561 & 139087 & 1023218\\
{16}  & 12733 & 77898 & 543240 & 4026672 
& 12492 & 76981 & 539875 & 4015237\\  
{32}  & 24734 & 153858 & 1081127 & 8030976
& 24409 & 152914 & 1080264 & 8017026 \\
\hline
\end{tabular}
\caption{Number of unknowns in the discrete system \eqref{eq: discrete general GNHEP} for different
  pipe lengths $(L)$ and mesh sizes $(h)$.}
\label{tab: unknowns vs L vs mesh_size}
\end{table}

\begin{table}[!ht]
  \centering
  \scriptsize
  \begin{tabular}{|l|llll|llll|}
\hline
& \multicolumn{4}{c|}{Homogeneous Dirichlet BCs} & \multicolumn{4}{c|}{Periodic BCs}   \\ 
\hline
\multicolumn{1}{|l|}{\backslashbox{$L$}{$h$}} & 0.5   & 0.25  & 0.125  & 0.0625 & 0.5   & 0.25  & 0.125  & 0.0625\\
\hline
{2}   & 1.40 & 3.52 & 21.49 & 643.29 
& 0.80 & 1.89 & 19.20 & 789.70 \\
{4}   & 0.84 & 8.44 & 56.75 & 1973.23 
& 0.74 & 5.16 & 74.31 & 2435.52\\
{16}  & 6.50 & 131.85 & 560.79 & 15513.43 
& 2.56 & 140.50 & 361.47 & 17561.25 \\ 
{32}  & 17.60 & 314.43 & 1340.82 & 30731.15
& 6.00 & 155.33 & 3129.13 & 33062.00 \\
\hline
\end{tabular}
  \caption{CPU time (in seconds) for obtaining the critical Reynolds number on pipes with length $L$
    and mesh size $h$ through solving \eqref{eq: discrete general GNHEP}.}
\label{tab: CPU time vs L vs mesh_size}
\end{table}

As illustrated in \Cref{tab: Re_c vs L vs mesh_size,tab: unknowns vs L vs mesh_size,tab: CPU time vs L vs mesh_size}
and as expected, refining the mesh improves the accuracy of the eigensolver but increases the number of unknowns and
the corresponding CPU time. From these results we can infer that mesh sizes $h = 0.25, 0.125$ provide a good balance
between accuracy and computational cost.

Examining the results from finer meshes in \Cref{tab: Re_c vs L vs mesh_size}, we observe that
the critical Reynolds number for homogeneous Dirichlet boundary conditions decreases monotonically
as the pipe length increases. In contrast, the behavior for periodic boundary conditions is
less clear, while there are minor fluctuations in the critical Reynolds number as the pipe length
increases, it does not demonstrate a consistent decrease. This behavior is further investigated in the next section.

\subsection{Critical Reynolds number as a function of pipe length}
\label{sec: Re vs L}
To investigate the relationship between the critical Reynolds number and pipe length,
we solved the eigenproblem for lengths ranging from $2\leq L < 20$ with fixed increments
of $0.05$. We shall consider both homogeneous Dirichlet and periodic boundary
conditions. To efficiently explore the range of lengths while maintaining good accuracy,
we utilized a coarser mesh size of $h = 0.25$.

\begin{figure}[h!]
\centering
\begin{subfigure}{.495\textwidth}
  \centering
  \includegraphics[width=\textwidth]{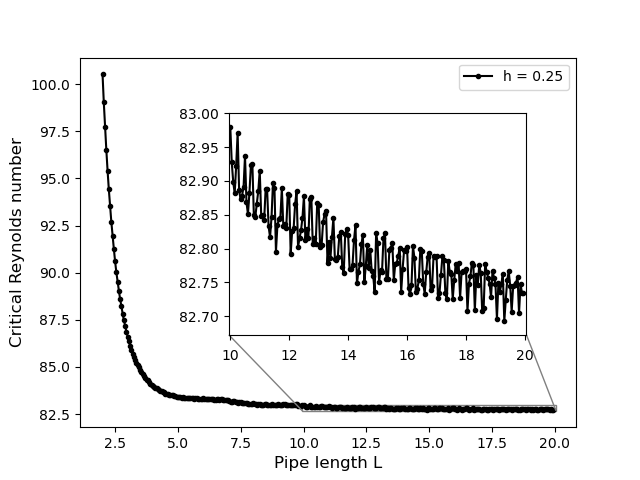}
  \caption{Homogeneous Dirichlet boundary conditions.}
  \label{fig: homo Rec of length L2-20}
\end{subfigure}
\hfill
\begin{subfigure}{.495\textwidth}
  \centering
  \includegraphics[width=\textwidth]{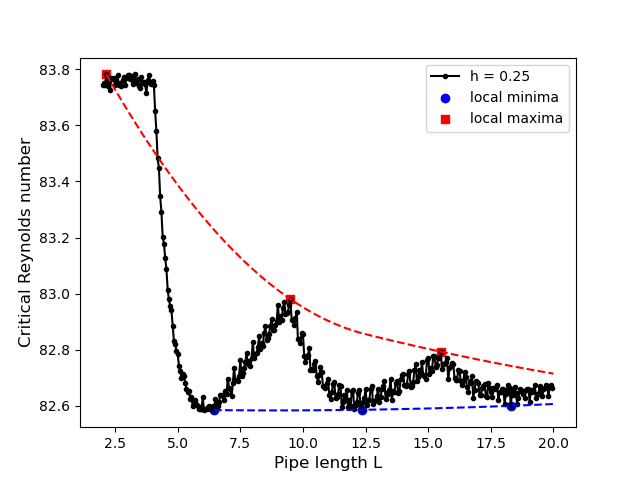}
  \caption{Periodic boundary conditions.}
  \label{fig: per Rec of length L2-20}
\end{subfigure}
\caption{Critical Reynolds number as a function of the pipe length for homogeneous Dirichlet (a) and periodic (b) boundary conditions. The critical Reynolds number is computed for pipe length in the range $[2, 20)$ with a length step of $0.05$. The mesh size is $h= 0.25$. Red and blue markers denote respectively the local maxima and minima. The oscillation shown in the inset are artifacts of discretization as discussed in \Cref{sec:osc}.
}
\label{fig: homo-per Rec of length}
\end{figure}

\Cref{fig: homo-per Rec of length} illustrates how the critical Reynolds number
varies\footnote{The high frequency oscillations present in the plots are due to discretization,
in particular, the geometric error in approximating the curved boundaries by affine elements.
The effect is further investigated in \Cref{sec:osc}.}
with pipe length for the eigenproblem under different boundary conditions. For
homogeneous boundary conditions, shown in \Cref{fig: homo Rec of length L2-20}, we observe
that the critical Reynolds number decreases with increasing pipe length. The decay is rapid
for shorter pipes but slows down as the length increases, suggesting a potential lower bound
for the function. In contrast, \Cref{fig: per Rec of length L2-20} reveals a periodic behavior
in the critical Reynolds number as a function of pipe length for periodic boundary conditions.
The local minima, indicated in blue in the figure, are reached at pipe length $L = 6.45, 12.35 \text{ and } 18.30$, with corresponding Reynolds number of $\RE_c = 82.58, 82.58 \text{ and } 82.59$. After a plateau region for shorter
pipe lengths $2 \leq L \leq 4.05$, the local maxima values decrease as pipe length increases, as shown by the red line in the figure. 

Our findings for periodic boundary conditions align with earlier results reported by Joseph and Carmi \cite{joseph1969stability}, who identified a critical Reynolds number of $\RE = 81.49$ for the z-periodic perturbation:
\begin{equation*}
    \vv(r,\theta,z) = \uu(r,\alpha,N) e^{(\alpha z +N\theta)}
\end{equation*}
for some function $\uu(r,\alpha,N)$, with $N=1$ and $\alpha = 1.07$. They obtained this result using polar
coordinates while rearranging equations \eqref{eq: parameter lambda} and \eqref{eq: perturbation NS}, concluding
that the eigenvalue is a function of $\alpha$ and $N$. They solved the resulting integral equations with
Runge-Kutta-Gill forward integration. The dimensionless wave number $\alpha = 1.07$ indicates that the perturbation is periodic in the $z-$direction with a period of 
\begin{equation}
    k = 2\pi/\alpha \approx 5.87.
\label{eq: JC periodicity}
\end{equation}

The rightward shift in our results, as presented in \Cref{fig: per Rec of length L2-20}, can be attributed to the
coarse mesh and the sampling of pipe lengths. Similarly, the higher value of the lower bound for the critical
Reynolds number is also due to the coarse mesh, as indicated in \Cref{tab: Re_c vs L vs mesh_size}. To
facilitate better comparisons with the results from Joseph and Carmi, we solved the eigenproblem for finer mesh
$h = 0.0625$ at specific lengths corresponding to local minima and maxima. The local minima occur at {$L = k, 2k, 3k$},
while the local maxima are found at {$L= \frac{3}{2}k, \frac52 k$ (with $k$ defined in \eqref{eq: JC periodicity})}.
We also examined the plateau region noted for shorter lengths,
selecting $L= 2.935$ {$\left(\frac12 k \right)$} as a representative value. \Cref{tab: per 0.0625 local max and min} displays the critical Reynolds
numbers associated with these points of interest. We observe that the local minima share the same value up to the second
decimal place, while the local maximum value decreases as the pipe length increases. This behavior aligns with observations
from \Cref{fig: per Rec of length L2-20}. 
\begin{table}[h!]
  \centering
  \scriptsize  
\begin{tabular}{ c |c c c c c c}
 \hline
 \multirow{2}{*}{length} & 2.935 & 5.87 & 8.805 & 11.74 & 14.675 & 17.61 \\
 & plateau & \text{local min} & \text{local max} & \text{local min} & \text{local max} & \text{local min} \\
$\RE_c$ & 82.93 & 81.58 & 81.97 & 81.58 & 81.74 & 81.58 \\ \hline
\end{tabular}
\caption{Pipe lengths $L$ and critical Reynolds numbers $\RE_c$ corresponding to the local minima and maxima,
  as well as the plateau region under periodic boundary conditions, for $2 \leq L \leq 20$, cf. \Cref{fig: per Rec of length L2-20}.
  The mesh size used is $h = 0.0625$.}
\label{tab: per 0.0625 local max and min}
\end{table}

To facilitate a comparison between the two boundary conditions, we also computed the critical Reynolds
number with homogeneous Dirichlet boundary conditions at the points corresponding to local minima and
maxima observed in the periodic case. These values are summarized in \Cref{tab: homo 0.0625 local max and min}.
As expected, the critical Reynolds number decreases with increasing pipe length, but the rate of decrease
slows down as the length continues to increase.
\begin{table}[h!]
  \centering
  \scriptsize
\begin{tabular}{ c | c c c c c c}
 \hline
 \text{length} & 2.935 & 5.87 & 8.805 & 11.74 & 14.675 & 17.61 \\  
 $\RE_c$ & 85.74 & 82.38 & 82.00 & 81.82 & 81.75 & 81.69 \\ \hline
\end{tabular}
\caption{Critical Reynolds numbers $\RE_c$ with homogeneous Dirichlet boundary conditions, relative
  to the local maxima and minima identified in the $\RE_c$ dependence with periodic boundary conditions,
  see \Cref{tab: per 0.0625 local max and min}.
  The mesh size used is $h = 0.0625$.}
\label{tab: homo 0.0625 local max and min}
\end{table}

\subsubsection{Numerical oscillation}\label{sec:osc}
\Cref{fig: homo-per Rec of length} reveals high frequency oscillations in the computed dependence of the critical Reynolds number on the pipe length. Here we determine
whether this behavior is a physical phenomenon or a numerical artifact. In particular,
the oscillations may be influenced by mesh size and may be amplified by geometric inaccuracies
caused by the polyhedral approximation of the cylinder.
To this end we compare our previous finite element discretization based on meshing the physical
domain with an alternative method of mapping approach. Here the computations are performed on a
reference (box) domain which is mapped onto the cylinder by a suitable bijective map. The finite
element mesh of the computational domain is there exact, however, the mapping introduces non-polynomial
integrands in the finite element forms. We refer to \ref{Appendix:Map from cylinder to box} for further
details.

\begin{figure}[h]
\centering
\begin{subfigure}{.49\textwidth}
  \centering
  \includegraphics[width=\textwidth]{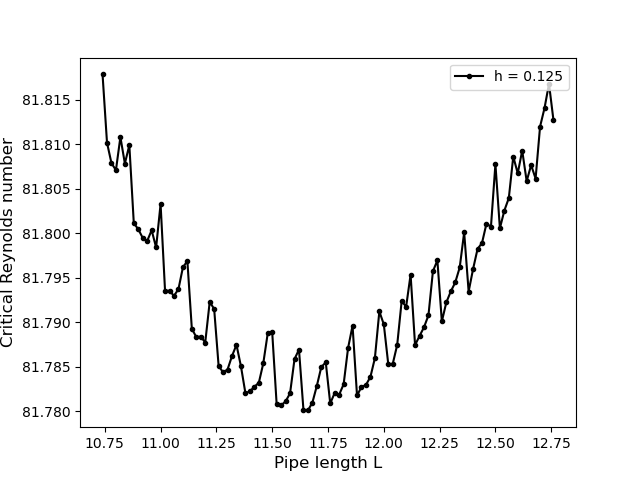}
  \caption{FEM with cylindrical domain}
  \label{fig: cylinder Rec of length L2-20}
\end{subfigure}
\hfill
\begin{subfigure}{.49\textwidth}
  \centering
  \includegraphics[width=\textwidth]{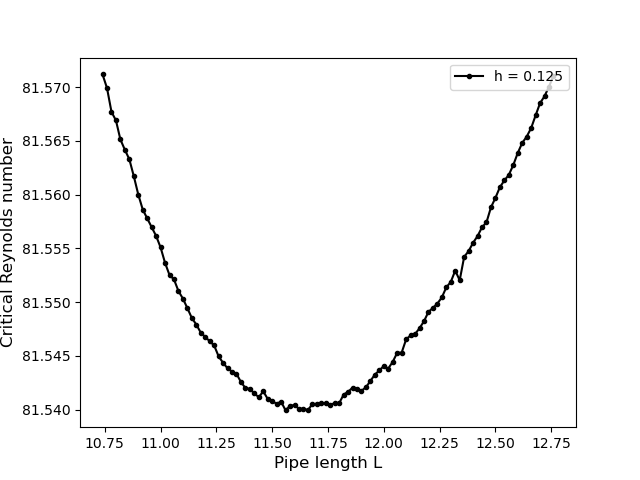}
  \caption{FEM with mapped domain}
  \label{fig: box Rec of length L10-12}
\end{subfigure}
\caption{Critical Reynolds number around a local minima at pipe length $L = 11.66$. Periodic boundary conditions and mesh size $h=0.125$ and pipe length $10.74 \leq L < 12.78 $ with length step of $0.02$. Solved in the cylinder's approximation (a) and in a mapped box domain (b).}
\label{fig: periodic Rec of length 10-12}
\end{figure}

The approaches are compared in \Cref{fig: periodic Rec of length 10-12} where interval of lengths
around $L=11.66$ is considered with periodic boundary conditions. We observe that with the new approach
the curve is visibly smoother than in \Cref{fig: cylinder Rec of length L2-20}, suggesting that the oscillations
in \Cref{fig: cylinder Rec of length L2-20} are a numerical artifact. 

\subsubsection{Characterization of the perturbations}
We shall characterize the perturbations in space by analyzing
their velocity, vorticity and resulting streamlines. For the sake
of easier comparison, as the first step in the analysis, we normalize the perturbations
such that ${\max_{\xx} \Vert \vv_{\lambda} (\xx) \Vert_2}=1$ for each perturbation. To obtain the vorticity,
we compute the $L^2$-projection of $\nabla\times\ww$ into the space of continuous piecewise
linear functions. The streamlines at given domains are seeded from points within a
unit sphere with center at the midpoint of the domain. We recall that $R=1$ in our
experiments. Finally, in the following we will visualize the vector fields, in particular
their magnitude, through $2D$ projections. These will be obtained either as planar slices
or by averaging over the angular direction. That is, for a function $f:\Omega\rightarrow\mathbb{R}$
(defined over the cylinder) its azimuth average $F:[0, R-\varepsilon]\times[0, L]\rightarrow\mathbb{R}$
\begin{equation}\label{eq:avg}
    F(r, z) = \frac{1}{2\pi} \int_{0}^{2\pi} f(r\sin \theta, r\cos\theta, z) \, \mathrm{d}\theta.
\end{equation}
The integral in the right-hand side of \eqref{eq:avg} is computed by trapezoidal rule
with $144$ partitions of the $[0, 2\pi)$ interval. We note that $\frac{2\pi}{0.0625} \approx 100$ segments.
Finally, $ \ \varepsilon = 10^{-2}$ is chosen to ensure that during the integration the quadrature
points are inside $\Omega$ (in the sense of finite precision arithmetic).

We analyze and compare the perturbations at various Reynolds numbers under both homogeneous
Dirichlet and periodic boundary conditions, focusing specifically on those associated with the
local minima and maxima, as reported in \Cref{tab: homo 0.0625 local max and min,tab: per 0.0625 local max and min}.
We also include the plateau region observed for shorter lengths, selecting {$L= \frac12 k$} as a representative value,
and solving for a finer mesh with size $h = 0.0625$. {Here $k$ is the period defined in \eqref{eq: JC periodicity}.}

\begin{figure}[h!]
\centering
\begin{subfigure}{.5\textwidth}
  \includegraphics[width=\textwidth]{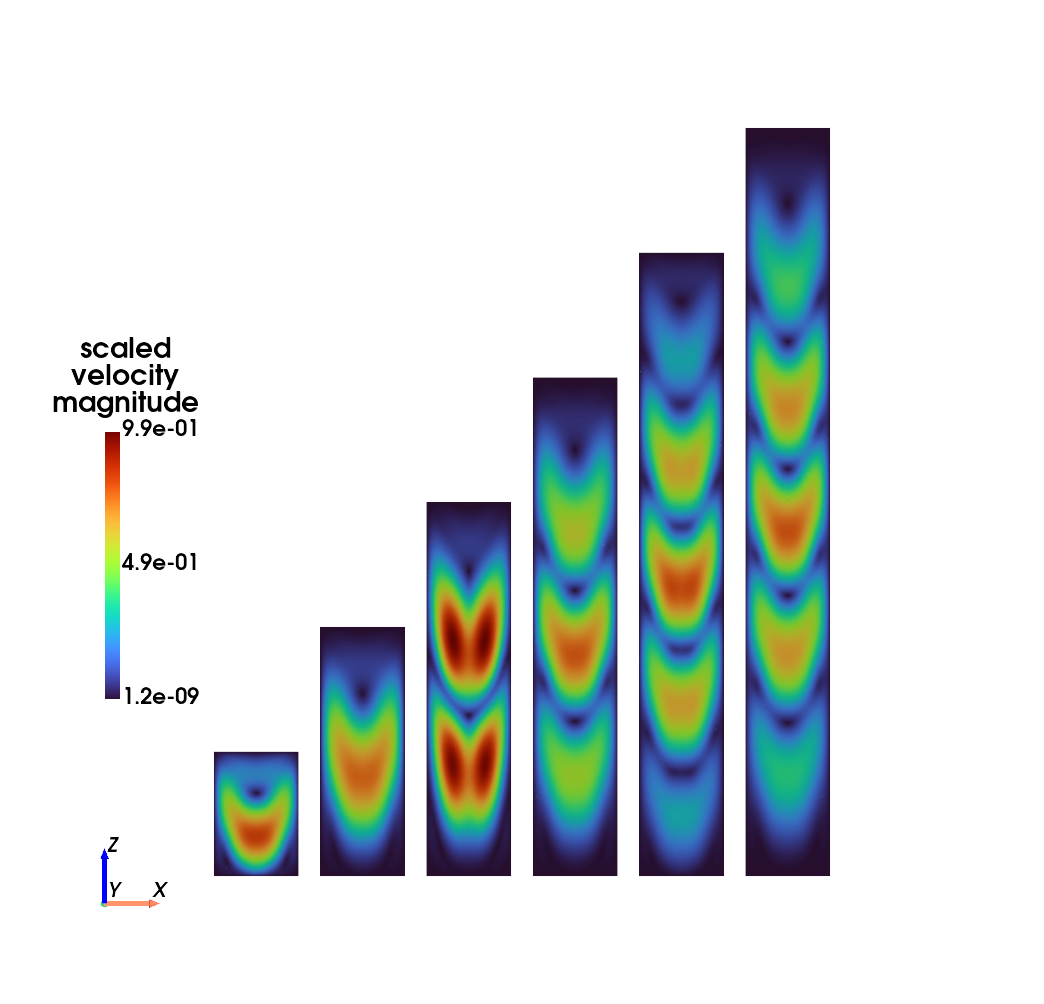}
  \caption{Homogeneous Dirichlet boundary conditions.}
\end{subfigure}%
\hfill
\begin{subfigure}{.5\textwidth}
  \includegraphics[width=\textwidth]{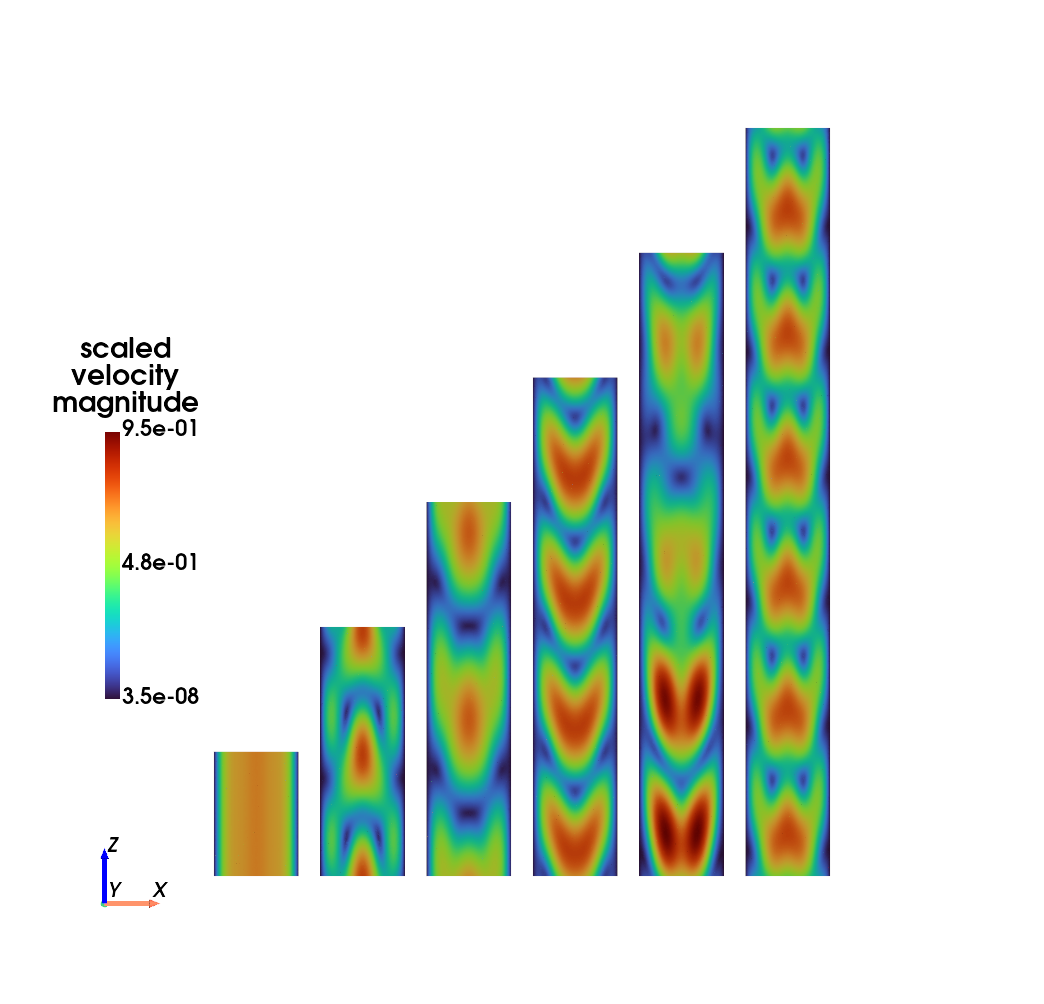}
  \caption{Periodic boundary conditions.}
\end{subfigure}
\caption{$xz$-Planar slices of the magnitude of the perturbations for various pipe lengths. From left to right, the pipe lengths are {$L= \frac12 k, k, \frac32 k, 2k, \frac52k, 3k$, with $k$ in \eqref{eq: JC periodicity}}. The first value corresponds to the plateau region, while the second, forth and the sixth values represent the local minima. The third and fifth values correspond to the local maxima. These lengths were identified in \Cref{tab: per 0.0625 local max and min}. All simulations were conducted using a mesh size of $h = 0.0625$.}
\label{fig:velocity_slice}
\end{figure}

\begin{figure}[h!]
\centering
\begin{subfigure}{.5\textwidth}
  \includegraphics[width=\textwidth]{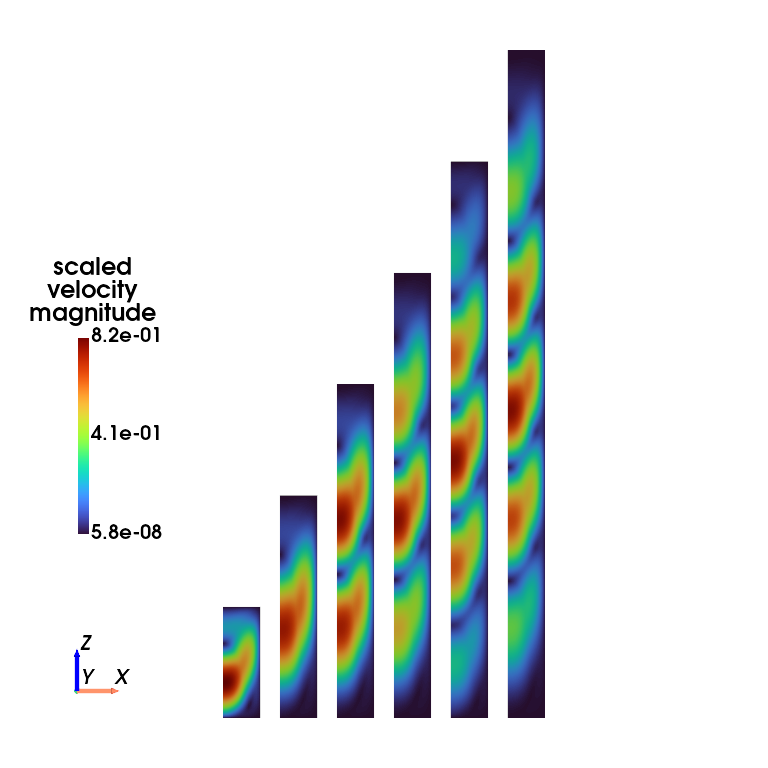}
  \caption{Homogeneous Dirichlet boundary conditions.}
  \label{fig:magnitude homo}
\end{subfigure}%
\hfill
\begin{subfigure}{.5\textwidth}
  \includegraphics[width=\textwidth]{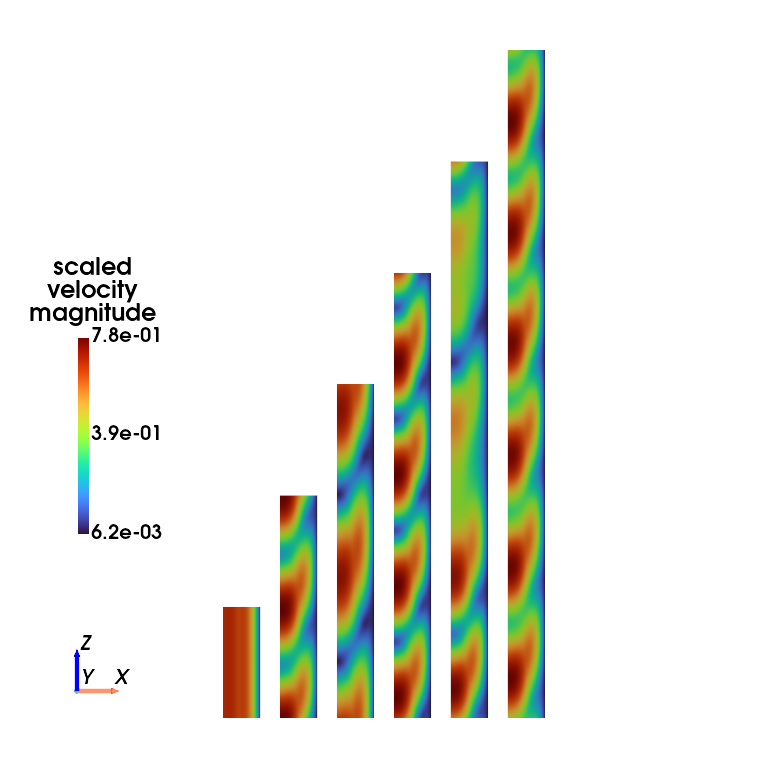}
  \caption{Periodic boundary conditions.}
  \label{fig:magnitude per}
\end{subfigure}
\caption{Azimuthal average of the magnitude of the perturbations for various pipe lengths. From left to right, the pipe lengths are {$L= \frac12 k, \, k, \, \frac32 k, \, 2k, \,\frac52 k, \, 3k$.} {The left-hand side in each panel corresponds to the centerline of the pipe.}} 
\label{fig:min_max_magnitude}
\end{figure}

\begin{figure}[h!]
\centering
\begin{subfigure}{.5\textwidth}
  \includegraphics[width=\textwidth]{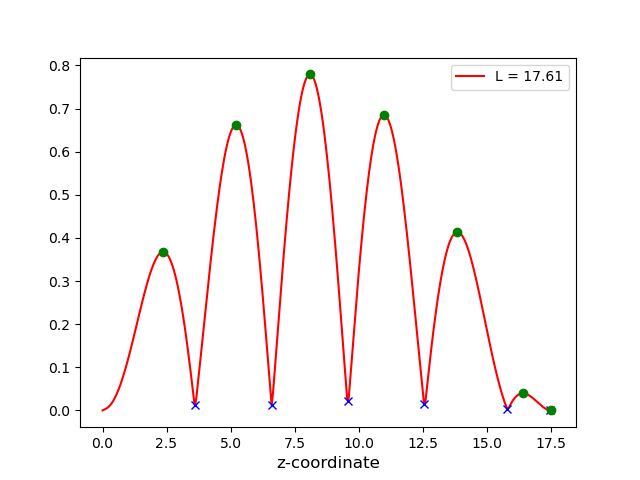}
  \caption{Homogeneous Dirichlet boundary conditions.}
  \label{fig: homo centerline L17}
\end{subfigure}%
\hfill
\begin{subfigure}{.5\textwidth}
  \includegraphics[width=\textwidth]{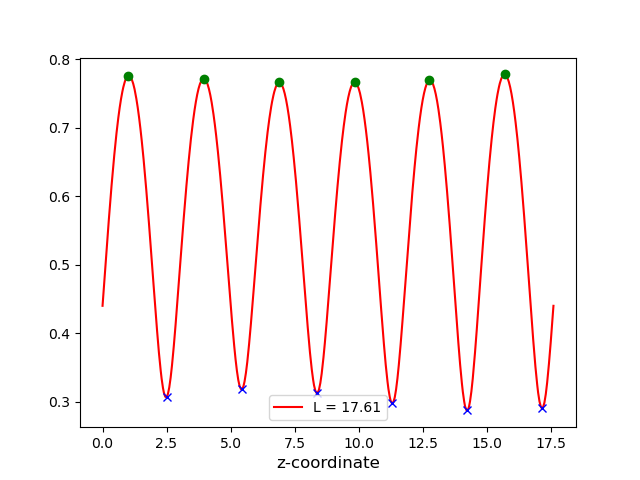}
  \caption{Periodic boundary conditions.}
  \label{fig:per centerline L17}
\end{subfigure}
\caption{Magnitude of the averaged velocity along the line passing through the points $(r=0, z=0)$ and $(r=0, z=L)$. The results are relative to the perturbations associated with a pipe length of {$L = 17.61 \, (3k)$} and a mesh size of $h=0.0625$.}
\label{fig:centerline L17}
\end{figure}

The analysis of the perturbation magnitudes presented in \Cref{fig:min_max_magnitude,fig:velocity_slice} reveals
a periodic nature in most cases, regardless of the boundary conditions. This periodicity is characterized
by a relatively constant interval among the pipe lengths. To evaluate the periodicity of the perturbation,
we focus on the magnitude of the velocity along the centerline, defined by the points $(r=0, z=0)$
and $(r=0, z=L)$.

For homogeneous Dirichlet boundary conditions, as shown in \Cref{fig:magnitude homo}, the perturbations
across different pipe lengths exhibit minimal variation, which is to be expected. The centerline consistently
shows a similar pattern across all pipe lengths. The peaks gradually increase in amplitude, reaching a maximum value of approximately $0.8$ at the midpoint of the pipe before gradually decreasing. \Cref{fig: homo centerline L17} illustrates the magnitude at the centerline for a pipe length of {$L=3k$ as an example with $k$ defined in \eqref{eq: JC periodicity}.}
The mean period between successive local maxima ranges from $2.33$ to $2.69$ for all pipe lengths, except for shorter pipes that exhibit only a single local maximum.

In contrast, the behavior for periodic boundary conditions is different. The perturbation in the shortest pipe, associated with the critical Reynolds number in the plateau region, exhibits a streamwise characteristic. Conversely, the perturbations associated with the local minima of critical Reynolds numbers {(specifically at $L= k, \, 2k, \, 3k$)} show periodic behavior. The magnitude along the centerline forms an oscillatory function with a nearly constant amplitude. The peaks and troughs show consistent heights across the pipes, with local minima values in the range $[0.1, 0.3]$ and local maxima values around $0.8$ occurring at regular intervals along the $x-$axis. The mean period for these oscillations is between $2.91$ and $2.94$. \Cref{fig:per centerline L17} showcases the magnitude at the centerline for the perturbation in the pipe of length {$L=3k$}. For the two perturbations in the lengths corresponding to the local maxima of the critical Reynolds number, occurring at {$L= \frac32 k, \, \frac52 k$}, the periodicity is less pronounced, or the mean period is longer.

%
\begin{figure}[h!]
\centering
\begin{subfigure}{.5\textwidth}
  \includegraphics[width=\textwidth]{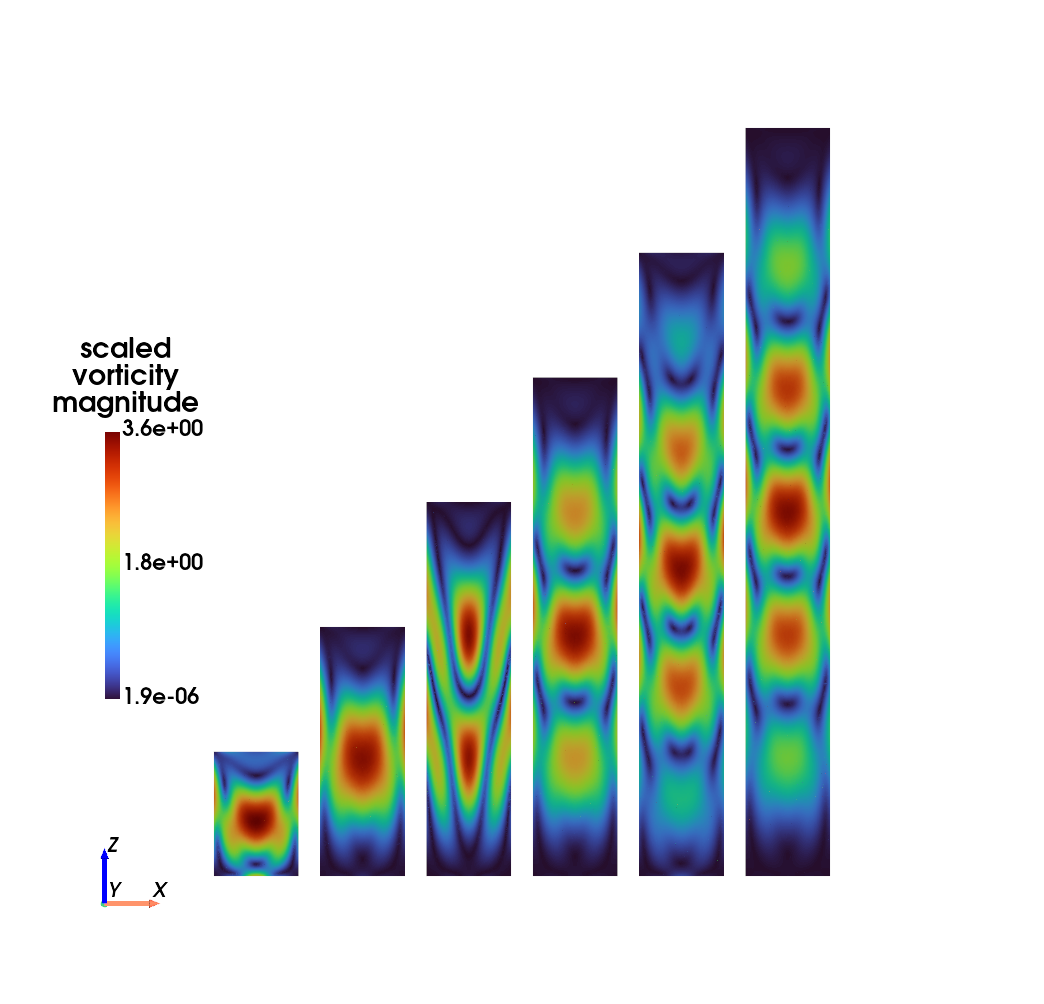}
  \caption{Homogeneous Dirichlet boundary conditions.}
\end{subfigure}%
\hfill
\begin{subfigure}{.5\textwidth}
  \includegraphics[width=\textwidth]{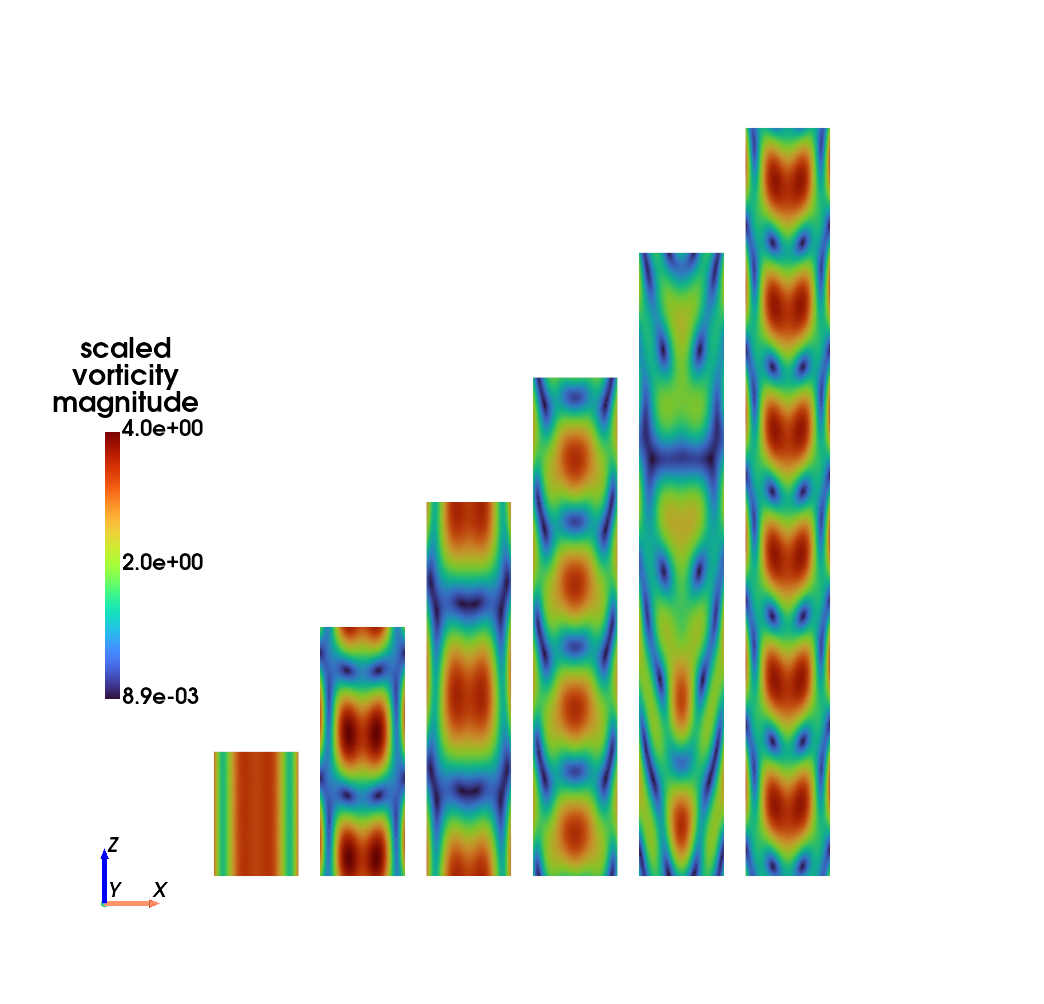}
  \caption{Periodic boundary conditions.}
\end{subfigure}
\caption{$xz$-Planar slices of the magnitude of the perturbations' vorticity for various pipe lengths. From left to right, the pipe lengths are {$L= \frac12 k, \, k, \, \frac32 k, \, 2k, \,\frac52 k, \, 3k$.}}
\label{fig:verticity_slice}
\end{figure}

\begin{figure}[h!]
\centering
\begin{subfigure}{.5\textwidth}
  \includegraphics[width=\textwidth]{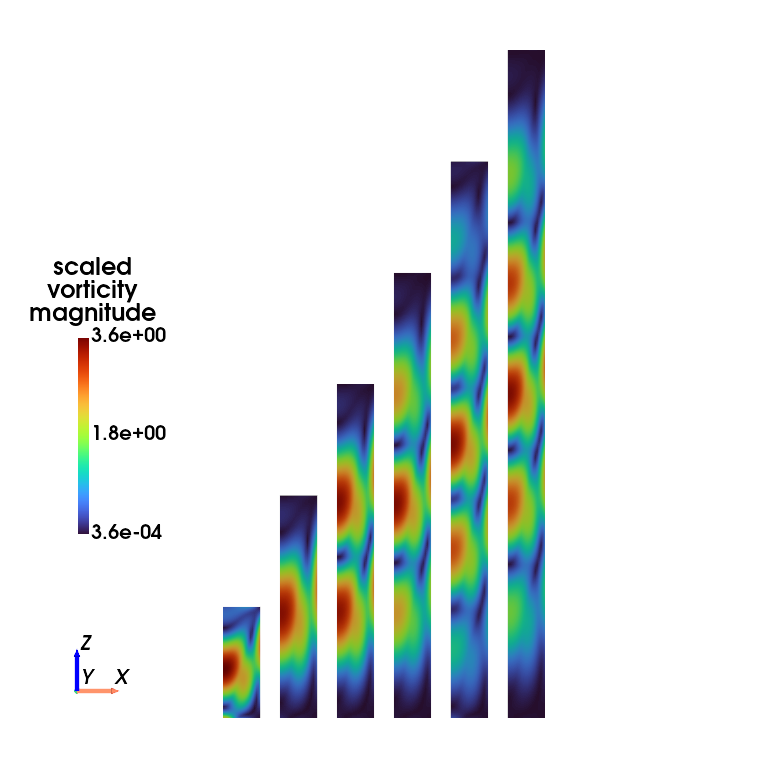}
  \caption{Homogeneous Dirichlet boundary conditions.}
  \label{fig:vorticity homo}
\end{subfigure}%
\hfill
\begin{subfigure}{.5\textwidth}
  \includegraphics[width=\textwidth]{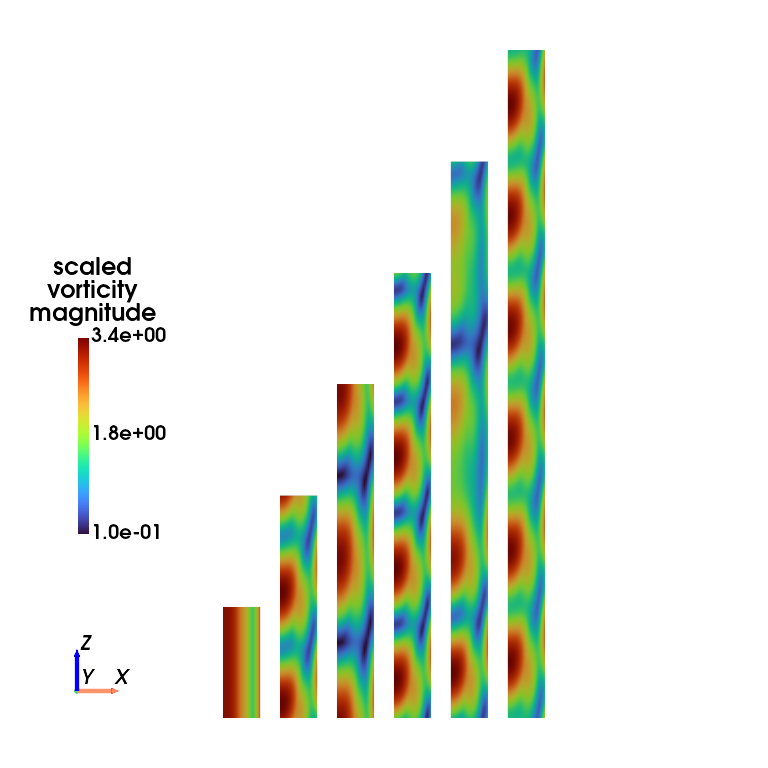}
  \caption{Periodic boundary conditions.}
  \label{fig:vorticity per}
\end{subfigure}
\caption{Azimuthal average of the magnitude of the perturbations' vorticity for various pipe lengths. From left to right, the pipe lengths are {$L= \frac12 k, \, k, \, \frac32 k, \, 2k, \,\frac52 k, \, 3k$.} {The left-hand side in each panel corresponds to the centerline of the pipe.}
}
\label{fig:min_max_vorticity}
\end{figure}

%
\begin{figure}[h!]
\centering
\begin{subfigure}{.5\textwidth}
  \includegraphics[width=\textwidth]{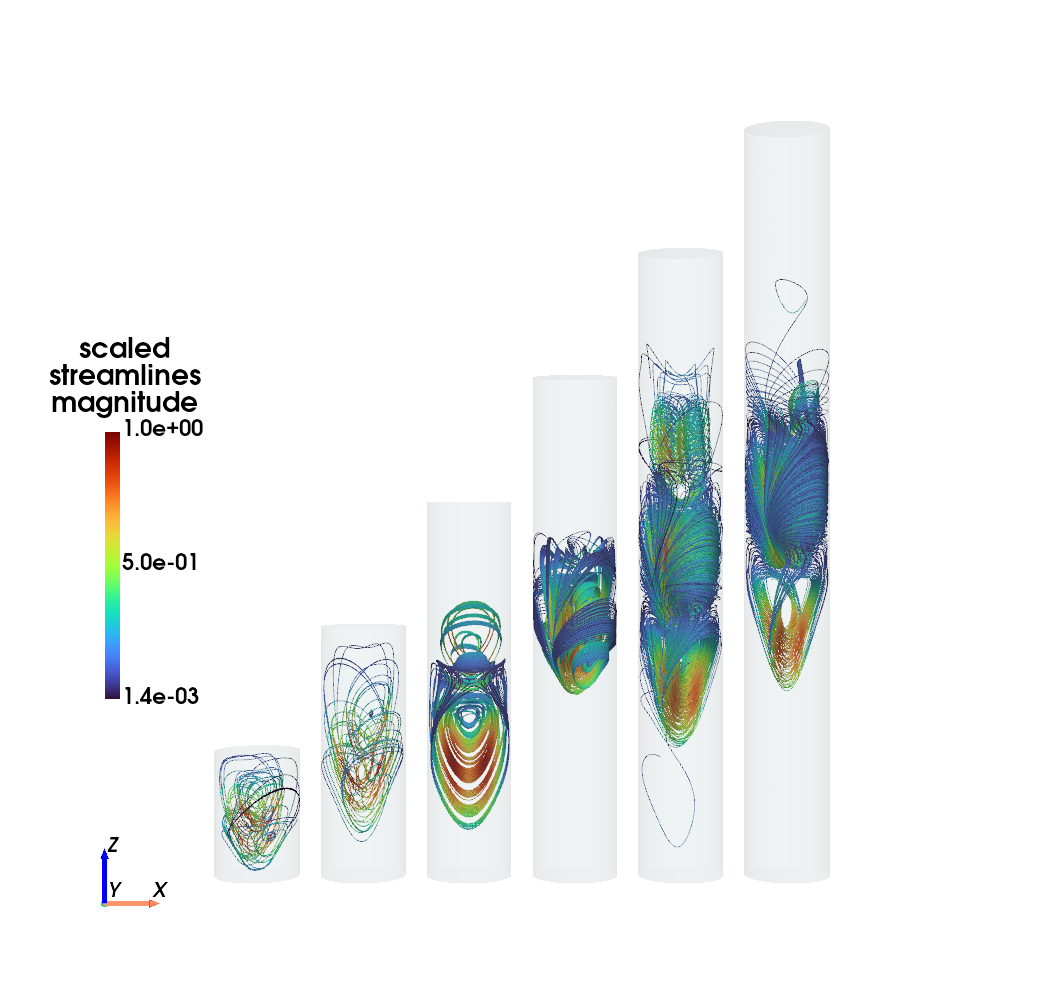}
  \caption{Homogeneous Dirichlet boundary conditions.}
  \label{fig:homo_streamlines_xz}
\end{subfigure}%
\hfill
\begin{subfigure}{.5\textwidth}
  \includegraphics[width=\textwidth]{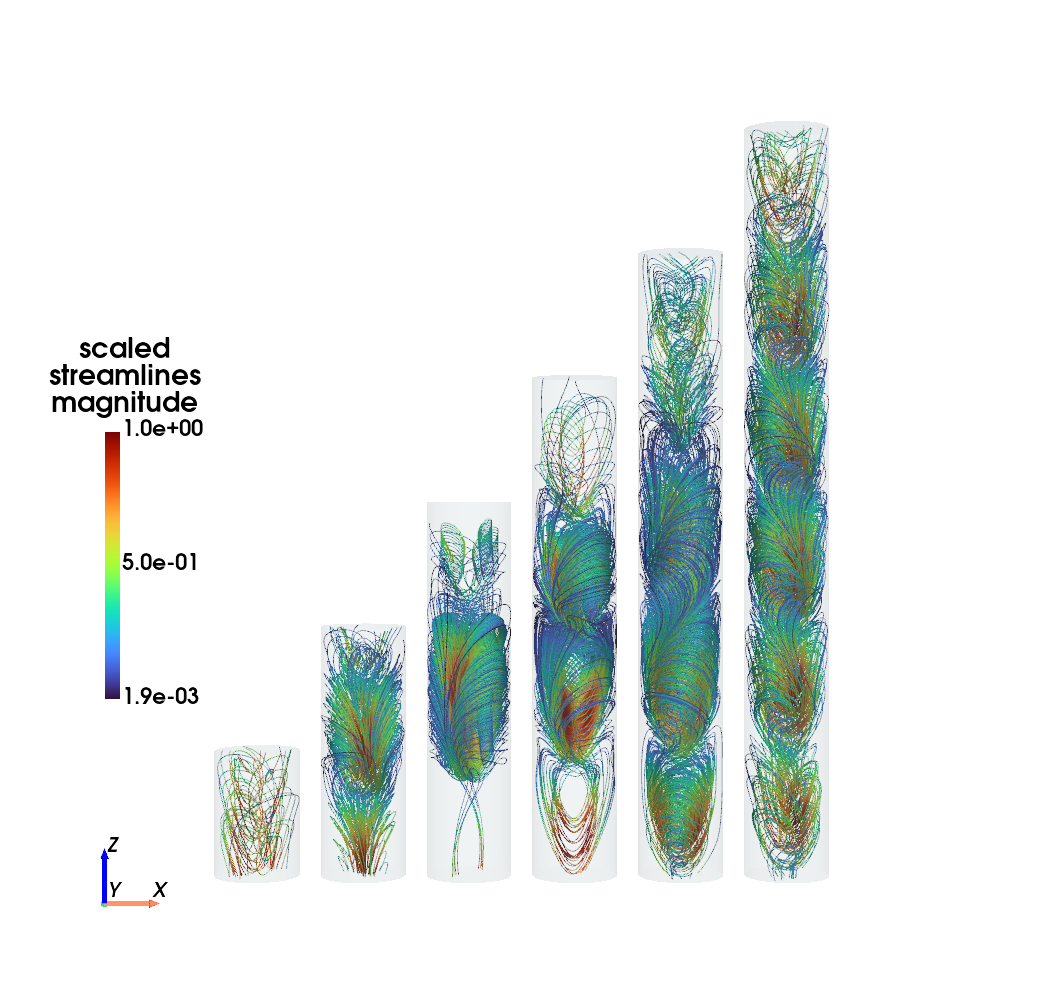}
  \caption{Periodic boundary conditions.}
  \label{fig:homo_streamlines_yz}
\end{subfigure}
\caption{Streamlines of the perturbations' vorticity for various pipe lengths from a $xz$-plane visualization. From left to right, the pipe lengths are {$L= \frac12 k, \, k, \, \frac32 k, \, 2k, \,\frac52 k, \, 3k$.}. All simulations were conducted using a mesh size of $h = 0.0625$.}
\label{fig:min_max_streamlines}
\end{figure}
%

Similar patterns in terms of symmetry and periodicity can be observed also in the magnitude of the vorticity, as shown in \Cref{fig:min_max_vorticity,fig:verticity_slice}. 
{The intricate spiral nature of the flow can be seen in the streamlines plots in \Cref{fig:min_max_streamlines}.}

\begin{figure}[h!]
\centering
\begin{subfigure}{.45\textwidth}
  \centering
  \includegraphics[width=\textwidth]{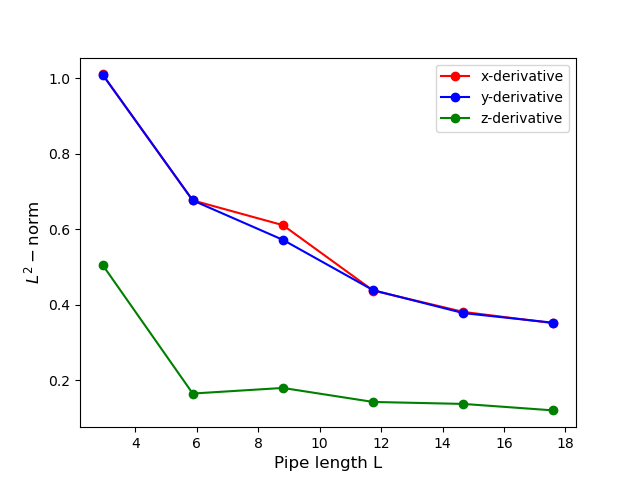}
  \caption{Homogeneous Dirichlet boundary conditions.}
  \label{fig: homo_L2_dv_dxdydz_maxmin}
\end{subfigure}
\hfill
\begin{subfigure}{.45\textwidth}
  \centering
  \includegraphics[width=\textwidth]{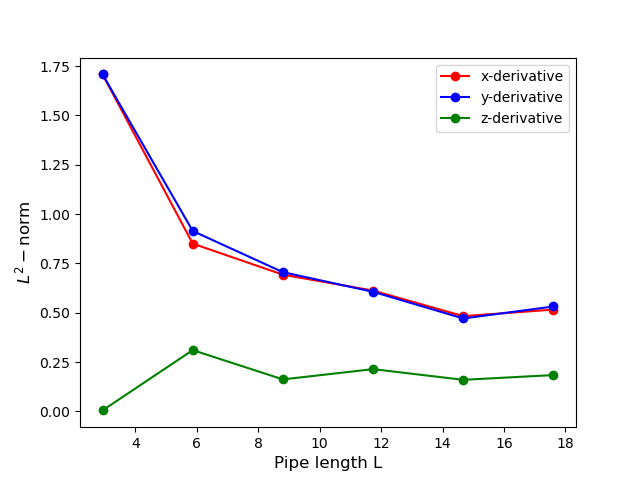}
  \caption{Periodic boundary conditions.}
  \label{ig: per_L2_dv_dxdydz_maxmin}
\end{subfigure}
\caption{$L^2-$norm of the partial derivatives of the perturbations associated to the pipe lengths {$L= \frac12 k, \, k, \, \frac32 k, \, 2k, \,\frac52 k, \, 3k$.}. All simulations were conducted using a mesh size of $h = 0.0625$.}
\label{fig: L2_derivatives}
\end{figure}
Finally, we compute the $L^2-$norm of the spatial derivatives of the perturbation,
i.e. $\lVert\tfrac{\partial \vv_\lambda}{\partial x_i}\rVert_2$. The results are displayed in
\Cref{fig: homo_L2_dv_dxdydz_maxmin} for the homogeneous Dirichlet boundary conditions and in \Cref{ig: per_L2_dv_dxdydz_maxmin} for the periodic boundary conditions. For both boundary conditions, Figures~\ref{fig: L2_derivatives} reveal that the perturbations depend on all three variables. However, since the $z-$derivative is smaller then the other two components, and the $x-$ and $y-$ derivatives exhibit similar and higher $L^2-$norm, it follows that the perturbations show a greater dependency on the spanwise components $x \text{ and } y$ than on the streamwise component $z$.   
   


\subsection{Dynamics of the perturbation over time}
To investigate the temporal evolution of perturbations in pipe flow, we numerically
compute the perturbed flow by solving the discrete form of the governing temporal-spatial
equations, as defined in equation \eqref{eq: perturbed ns system}, with initial conditions
defined in \eqref{eq: perturbed scaled ICs} where the scaling factor for the perturbation
is fixed to $1/4$. In order to achieve a good balance between accuracy and computational cost,
we solve the dynamical problem in a pipe of length $L=4$. To ensure that the perturbation reaches
the end of the pipe, we run the simulations until $T = 10$. Since equations \eqref{eq: perturbed ns system} solve for the perturbed flow $\ww$, and the steady-state solution $\mathbf{u}_{\text{HP}}$, defined in \eqref{eq: HP flow}, does not depend on time, the evolution on time of the perturbation $\vv$ is post-computed as
\begin{equation}
    \vv(t, \xx) = \ww(t, \xx) - \mathbf{u}_{\text{HP}}(\xx) \quad \forall t \in [0, T] \text{ and } \forall \xx \in \Omega. 
\end{equation}
We shall solve the {dynamical problem} only using Dirichlet boundary conditions for two primary reasons. First,
dynamical behavior of the perturbation does not necessarily stay periodic. Second, imposing periodic
boundary conditions does not allow for energy to enter
the system, leading the perturbation and the flow to decay due to viscous dissipation. 

Since our analysis is based on the kinetic energy theory from Reynolds and Orr \cite{ref:SerrinStabilityReOrr}, we
compute the $L^2-$norm of the perturbation and its time derivative, to observe their evolution over time.
We shall also separately consider the streamwise- ($\mathbf{e}_z\vv\cdot\mathbf{e}_z$) and spanwise-
($\vv-(\vv\cdot\mathbf{e}_z)\cdot\mathbf{e}_z$) components and evaluate their respective $L^2$ norms.

In the following we first investigate the role of the mesh size on the evolution in time of the perturbation, by solving the dynamical problem \eqref{eq: perturbed ns system} for different mesh sizes and their relative critical Reynolds numbers.
Then, having established a mesh size with a good accuracy, we examine the behavior of the perturbation at Reynolds
numbers equal to and exceeding the critical value.

\subsubsection{Effects of mesh size on the temporal evolution of the perturbation}
Before running the simulations to assess the impact of mesh size on the convergence of the solution,
we first analyze the computational cost associated with the numerical methods Crank-Nicholson (CN) and
Crank-Nicholson Adams-Bashforth (CN-AB), cf. \Cref{sec:dyndiscrete}.
 The simulations were conducted on an Apple MacBook Pro equipped with the Apple M1 Pro chip.
\Cref{tab: CPU time dynamics} presents the CPU time (in seconds) for mesh sizes $h=0.5$ and $0.25$ for
the two numerical methods. The simulations were performed over the time interval $[0,10]$ with a uniform time step of $\Delta t = 0.01$. 
\begin{table}[h!]
  \centering
  \scriptsize
\begin{tabular}{ c c c c }
 \hline
 $h$  & \# unknowns & CN CPU time (s) & CN-AB CPU time (s)\\ 
\hline 
$0.5 $ & $ 4174$ & $ 201.07$ & $ 125.9$ \\  
$0.25 $ & $ 21434$ & $2170.71$ & $ 1403.35 $\\
\hline
\end{tabular}
\caption{Comparison of CPU time (in seconds) for the dynamical problem solved using Crank-Nicholson (CN) and Crank-Nicholson Adams-Bashforth (CN-AB) methods, for the coarser mesh sizes $h=0.5 \text{ and } 0.25$. The number of degrees of freedom (dofs) associated with each mesh size is also reported.}
\label{tab: CPU time dynamics}
\end{table}
Based on the CPU time shown in \Cref{tab: CPU time dynamics},  we establish the following approach for the simulations discussed later: for the coarser meshes, $h= 0.5 \text{ and } 0.25$, we employ Crank-Nicholson formulation, while
for the finer meshes $h= 0.125 \text{ and } 0.0625$, we utilize the Crank-Nicholson Adams-Bashforth formulation, to
balance between accuracy and computational cost.
\begin{figure}[h!]
\centering
\begin{subfigure}{.8\textwidth}
  \centering
  \includegraphics[width=\textwidth]{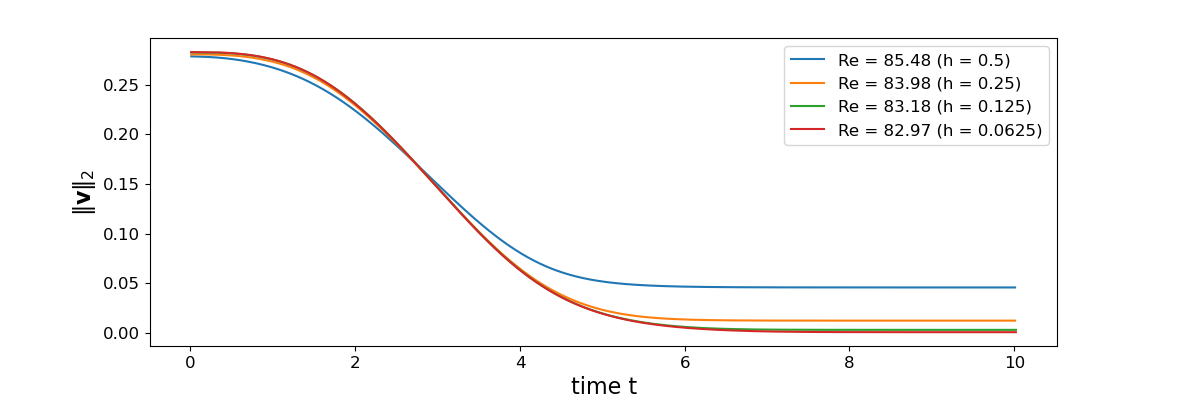}
  \caption{$L^2-$norm of the perturbation.}
  \label{fig:L2norm_perturbation_Rec}
\end{subfigure}%
\hfill
\begin{subfigure}{.8\textwidth}
  \centering
  \includegraphics[width=\textwidth]{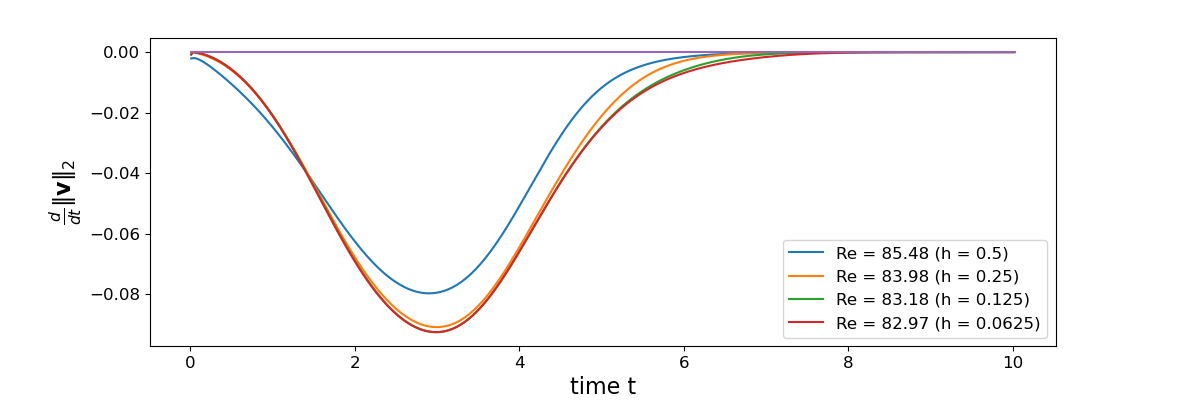}
  \caption{Time derivative of the $L^2-$norm of the perturbation.}
  \label{fig:dt_L2norm_perturbation_Rec}
\end{subfigure}
\caption{Temporal evolution of the $L^2$-norm of the perturbation (a) and its time derivative (b) for varying mesh sizes, along with their corresponding critical Reynolds numbers. The purple line in the bottom figure indicates the value of zero.}
\label{fig:L2norm_Rec}
\end{figure}
%
\begin{table}[h!]
  \centering
  \scriptsize
\begin{tabular}{ c c c c c}
 \hline
 $h$ & $0.5 $& $0.25$ & $0.125$ & $0.0625$ \\ 
\hline 
$\Vert \vv(\xx, t=10) \Vert_2 $ & $4.58 \times 10^{-2}$ & $1.24 \times 10^{-2}$ & $3.11 \times  10^{-3}$ & $7.77 \times 10^{-4}$ \\  
$\tfrac{\mathrm{d}\Vert \vv(\xx, t=10) \Vert_2}{\mathrm{d}t}$ & $-5.69 \times 10^{-7}$ & $-3.61 \times 10^{-8}$ & $-1.02 \times 10^{-6}$ & $-3.83 \times 10^{-6}$\\ \hline
\end{tabular}
\caption{
  Characteristics of the evolved scaled perturbation for length $L=4$ at time $t=10$.
  In agreement with the expected dissipation of the perturbation, convergence of the the
  $L^2$ norm of the energy with $h$ towards 0 is observed.}
\label{tab: L2norm final time}
\end{table}

\Cref{fig:L2norm_Rec} illustrates the temporal evolution of the $L^2-$norm of the perturbations (top) and its time derivative (bottom), relative to the critical Reynolds numbers for different mesh sizes, namely $h = 0.5, \, 0.25, \, 0.125, \, 0.0625$. To facilitate a comparison of results across mesh sizes, we present the values of the $L^2-$norm and its time derivative at the final computational time $t=10$ in \Cref{tab: L2norm final time}. The initial perturbation is derived from the scaled eigenmode corresponding to the respective mesh size and critical Reynolds number. 

The results across all mesh sizes align with the theories of Reynolds and Orr, showing that the $L^2-$norm of the perturbation decreases over time from its initial value, as depicted in the top plot of \Cref{fig:L2norm_Rec}. This trend is more pronounced in the bottom plot, where the time derivative of the $L^2-$norm remains negative, initially decreasing from zero before eventually increasing again as it converges towards zero. The upper plot in \Cref{fig:L2norm_Rec} and the data in \Cref{tab: L2norm final time} indicate that as time progresses and the mesh is refined, the $L^2-$norm of the perturbation approaches zero more closely, reflecting improved convergence. From the theoretical perspective, we expect the perturbation to dissipate over time, which implies that its $L^2-$norm will decrease to zero. This dissipation also means that the perturbed flow converges to the steady-state solution described by
\eqref{eq: HP flow}.

\begin{figure*}[h!]
\centering
\begin{subfigure}{.75\textwidth}
  \centering
  \includegraphics[width=\textwidth]{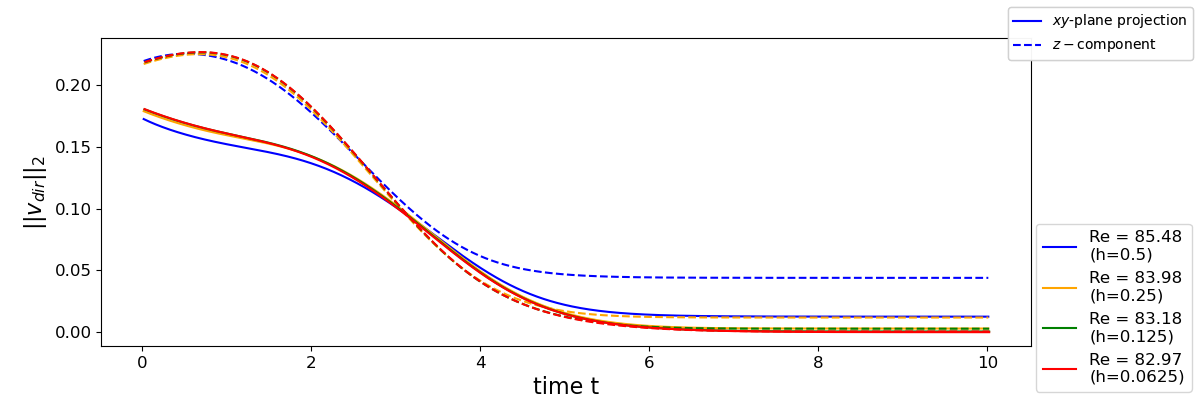}
  \caption{$L^2-$norm of the streamwise and spanwise components of the perturbation.}
  \label{fig:dL2norm_perturbation_Rec}
\end{subfigure}%
\hfill
\begin{subfigure}{.75\textwidth}
  \centering
  \includegraphics[width=\textwidth]{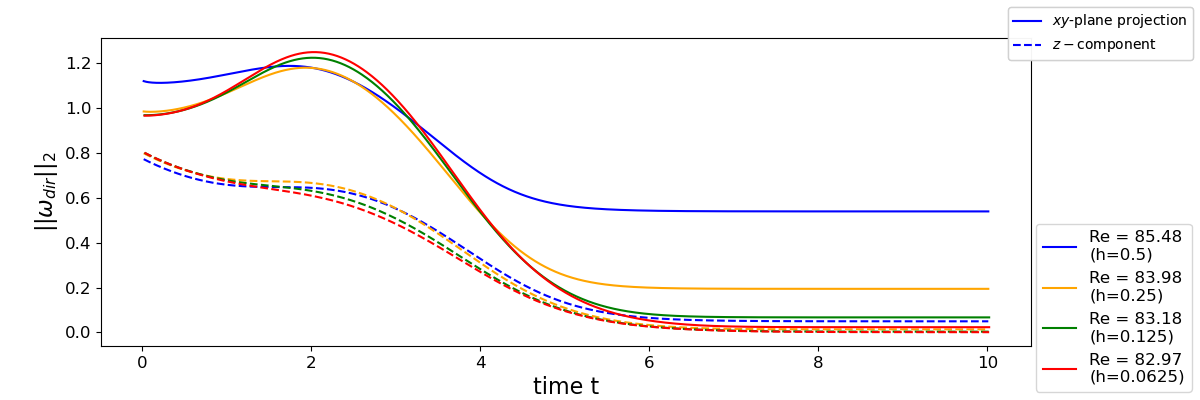}
  \caption{$L^2-$norm of the streamwise and spanwise components of the perturbation's vorticity.}
  \label{fig:dL2norm_vorticity_Rec}
\end{subfigure}
\caption{Directional plots illustrating the evolution over time of the $L^2-$norm of the streamwise ($z-$component) and spanwise ($xy-$plane projection) components of the perturbation and their vorticity for different mesh sizes: $h = 0.5, \, 0.25,\, 0.125, \, 0.0625$. Each solution corresponds to its associated critical Reynolds number.}
\label{fig:L2directional Rec}
\end{figure*}
The convergence of the solution with respect to mesh size is further supported by \Cref{fig:L2directional Rec}, which presents the $L^2-$norm of both the streamwise ($z$-component) and spanwise ($xy-$plane projection) components of the perturbations and and their vorticities over time. For coarser meshes, the streamwise component exhibits a larger convergence error, even though both components should ultimately converge to zero. \Cref{fig:dL2norm_vorticity_Rec} shows that the vorticity also converges to zero as the mesh is refined. Furthermore, the $L^2-$norm of the components behaves as anticipated: the vorticity in the $z-$direction indicates rotation around the $z-$axis that exists within the $xy-$plane, while the vorticity in the $xy-$plane corresponds to rotation around the axes within that plane. Consequently, the $L^2-$norms of the vorticity in the $xy-$plane are greater than those for the $z-$component, which aligns with the behavior of the $L^2-$norm of the perturbation, where the $xy-$plane projections are smaller than the $z-$component.

\subsubsection{Effects of the Reynolds number on the temporal evolution of the perturbation}
\label{subsubsection: dynamics numerics}
In this section we shall investigate temporal stability of the
perturbed flow for $\RE\geq\RE_c$.
To provide improved accuracy over the previous section
(where $h=0.125$ was satisfactory) we here consider 
the finest mesh, i.e. $h= 0.0625$.

%
\begin{figure}[h!]
\centering
\begin{subfigure}{.8\textwidth}
  \centering
  \includegraphics[width=\textwidth]{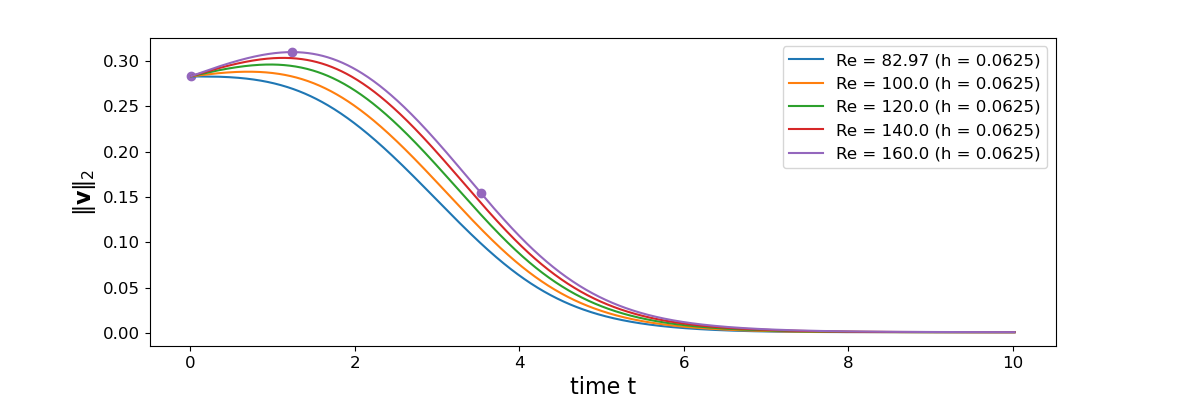}
  \caption{$L^2-$norm of the perturbation.}
  \label{fig:L2norm_perturbation_Rebig}
\end{subfigure}%
\hfill
\begin{subfigure}{.8\textwidth}
  \centering
  \includegraphics[width=\textwidth]{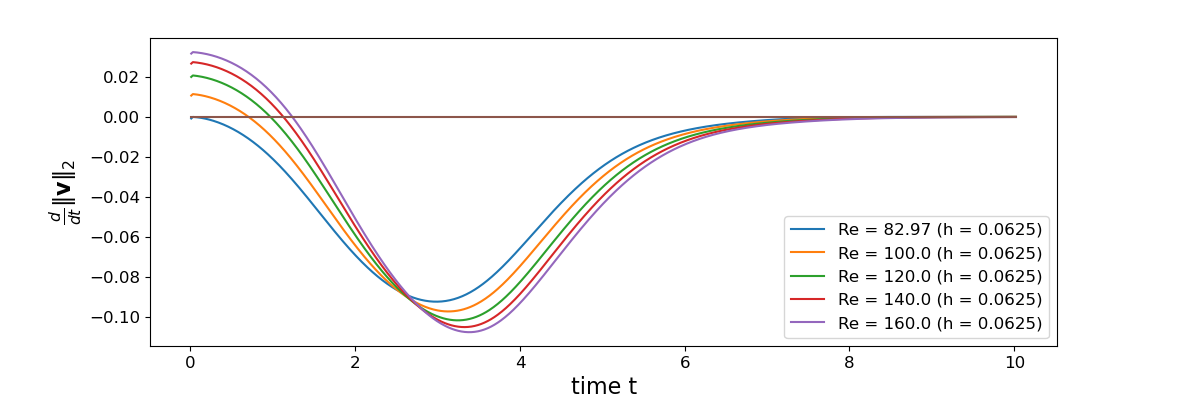}
  \caption{Time derivative of the $L^2-$norm of the perturbation.}
  \label{fig:dt_L2norm_perturbation_Rebig}
\end{subfigure}
\caption{Temporal evolution of the $L^2$-norm of the perturbation (a) and its time derivative (b) for Reynolds number equal to and greater than the critical one. Simulations were conducted using the finest mesh with size $h = 0.0625$. In figure (a) for $Re = 160$, the dots indicate the initial time $(t=0)$, the time at which the $L^2$-norm of the perturbation reaches its maximum $(t= 1.24)$, and a time midway through the decreasing phase $(t= 3.53)$.}
\label{fig:L2norm_Rebig}
\end{figure}
%
\begin{table}[h!]
\centering
\begin{tabular}{ c c c c }
 \hline
 Re  & $t_{\max}$ & $\Vert \vv(\xx, t_{\max}) \Vert_{2} $ & $t_{\text{decay}}$ \\ 
\hline 
$100 $ & $ 0.61$ & $ 0.288$ & $ 4.38$  \\  
$120 $ & $ 0.97$ & $ 0.296$ & $ 4.54$  \\ 
$140 $ & $ 1.13$ & $ 0.303$ & $ 4.67$  \\ 
$160 $ & $ 1.24$ & $ 0.309$ & $ 4.76$  \\
\hline
\end{tabular}
\caption{Summary of the maximum time $t_{\max}$ at which the $L^2-$norm of the perturbation reaches its peak,
  the maximum value achieved, and the decay time $t_{\text{decay}}$ defined as the time when the $L^2-$norm falls
  below a threshold value $0.05$. These results correspond to the temporal evolution of the perturbation for
  Reynolds numbers greater than the critical value, using the finest mesh size $h=0.0625$.}
\label{tab: t max and decay}
\end{table}
\Cref{fig:L2norm_Rebig} illustrates the temporal evolution of the $L^2-$norms of the perturbation (top) and its time derivative (bottom) for Reynolds numbers equal to or greater than the critical value. The data shown in this figure are derived from the same scaled initial perturbation but computed for different Reynolds numbers, specifically $\RE = 82.97$, which is the critical Reynolds number, and higher values $\RE = 100, 120, 140 \text{ and } 160$. Including the critical Reynolds number allows for a direct comparison. As seen in \Cref{fig:L2norm_Rebig}, when solving the equations with the critical Reynolds number, the $L^2-$norm of the perturbation decreases from its initial value (showing a negative derivative at $t=0$). In contrast, when the same equations are solved with Reynolds numbers above the critical value, the $L^2-$norm initially increases, reaching a maximum before eventually decreasing (showing a positive derivative at the initial time), which aligns with the theory established by Reynolds and Orr \cite{ref:SerrinStabilityReOrr}.

The behavior observed in \Cref{fig:L2norm_Rebig} and summarized in \Cref{tab: t max and decay} for Reynolds numbers greater than the critical one indicates a monotonic relationship between the $L^2-$norm of the perturbation and the Reynolds number. Specifically, as the Reynolds number increases, both the maximum value of the $L^2-$norm and the time at which it occurs increase as well. Additionally, the decay time, defined as the first time when the $L^2-$norm falls below $0.05$, also increases with higher Reynolds numbers.

To conduct a more comprehensive analysis of the temporal evolution of the perturbation, we also compute the $L^2-$norm of the streamwise ($z-$component) and spanwise ($xy-$plane projection) components of both the perturbation and its vorticity.
\begin{figure*}[h!]
\centering
\begin{subfigure}{.75\textwidth}
  \centering
  \includegraphics[width=\textwidth]{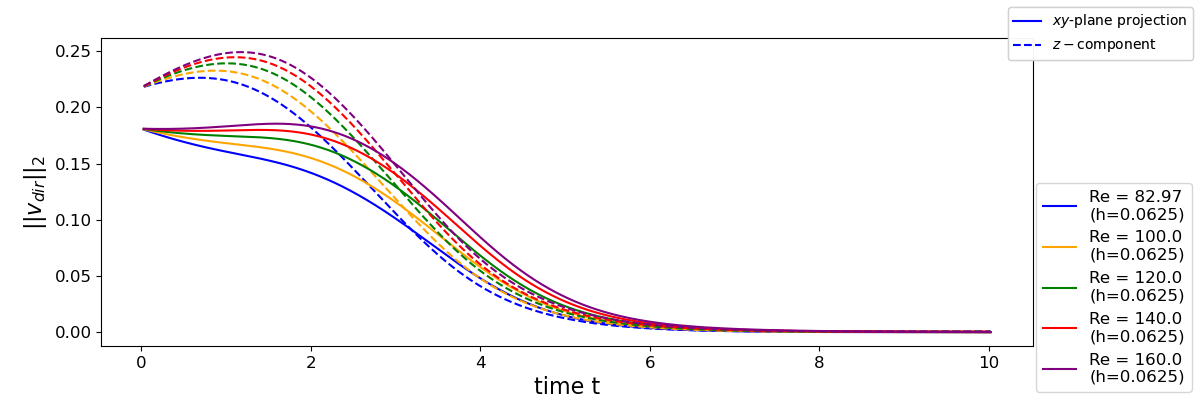}
  \caption{$L^2-$norm of the streamwise and spanwise components of the perturbation.}
  \label{fig:dL2norm_perturbation_Rebig}
\end{subfigure}%
\hfill
\begin{subfigure}{.75\textwidth}
  \centering
  \includegraphics[width=\textwidth]{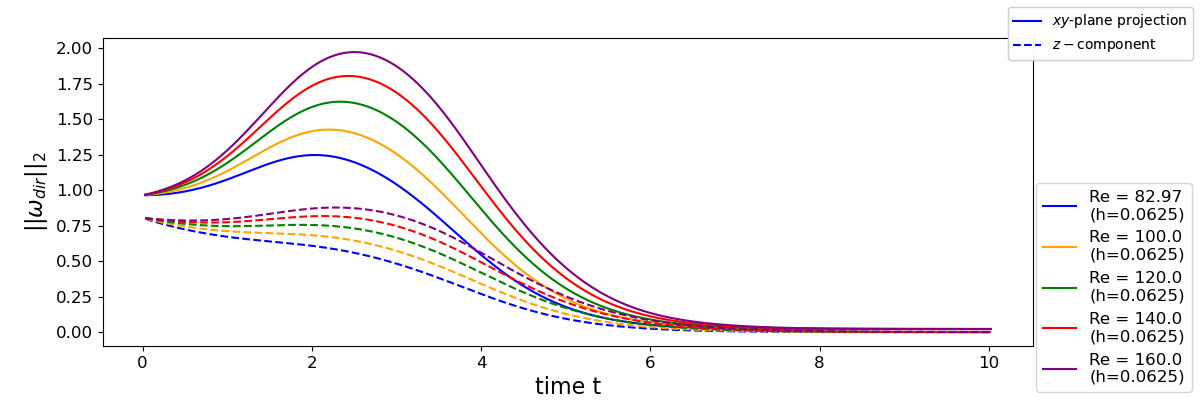}
  \caption{$L^2-$norm of the streamwise and spanwise components of the perturbation's vorticity.}
  \label{fig:dL2norm_vorticity_Rebig}
\end{subfigure}
\caption{Directional plots illustrating the evolution over time of the $L^2-$norm of the streamwise ($z-$component) and spanwise ($xy-$plane projection) components of the perturbation and its vorticity for Reynolds numbers equal to the critical value $\RE_c = 82.97$ and larger $(Re = 100, \, 120, \, 140, \, 160)$. The mesh size is fixed at $h = 0.0625$.}
\label{fig:dplot_Rebig}
\end{figure*} 

Interestingly, the $L^2-$norm of the $z-$component of the perturbation exhibits an initial increase regardless of whether the Reynolds number is equal to or exceeds the critical value, as shown in \Cref{fig:dL2norm_perturbation_Rebig}. However, considering the results in \Cref{fig:L2norm_Rebig,fig:dL2norm_perturbation_Rebig}, we can infer that for the Reynolds number at the critical level, the $L^2-$norm of the $xy-$projection of the perturbation decreases initially at a faster rate than the increase of the $L^2-$norm of the $z-$component, resulting in an overall decrease in the $L^2-$norm of the perturbation. For the cases with Reynolds numbers greater than the critical value, the $L^2-$norm of the $xy-$projection only begins to increase once the Reynolds number exceeds a certain threshold (around $\RE = 140$). For all Reynolds numbers greater than the critical value, the $L^2-$norm of the $z-$direction component initially increases at a faster rate than the changes in the spanwise component, leading to an overall increase in the $L^2-$norm of the perturbation. This indicates that the main contributor to this initial instability is the streamwise component of the perturbation.

These observations are further supported by the $L^2-$norm of the vorticity in both the $z-$direction and $xy-$plane projection for both critical and higher Reynolds numbers, as shown in \Cref{fig:dL2norm_vorticity_Rebig}. As previously mentioned in section~\ref{subsubsection: dynamics numerics}, the $L^2$-norms of the streamwise velocity component behave similarly to those of the spanwise vorticity, while the spanwise velocity component resembles the streamwise vorticity. \Cref{fig:dL2norm_vorticity_Rebig} shows that the $L^2-$norm of the vorticity in the $xy-$plane projection initially increases for all Reynolds numbers. In contrast, the $L^2-$norm of the $z-$component of the vorticity decreases for the critical Reynolds number, and either decreases or shows slight increases for Reynolds numbers greater than the critical value.


\begin{figure}[h!]
\centering
  \includegraphics[width=\textwidth]{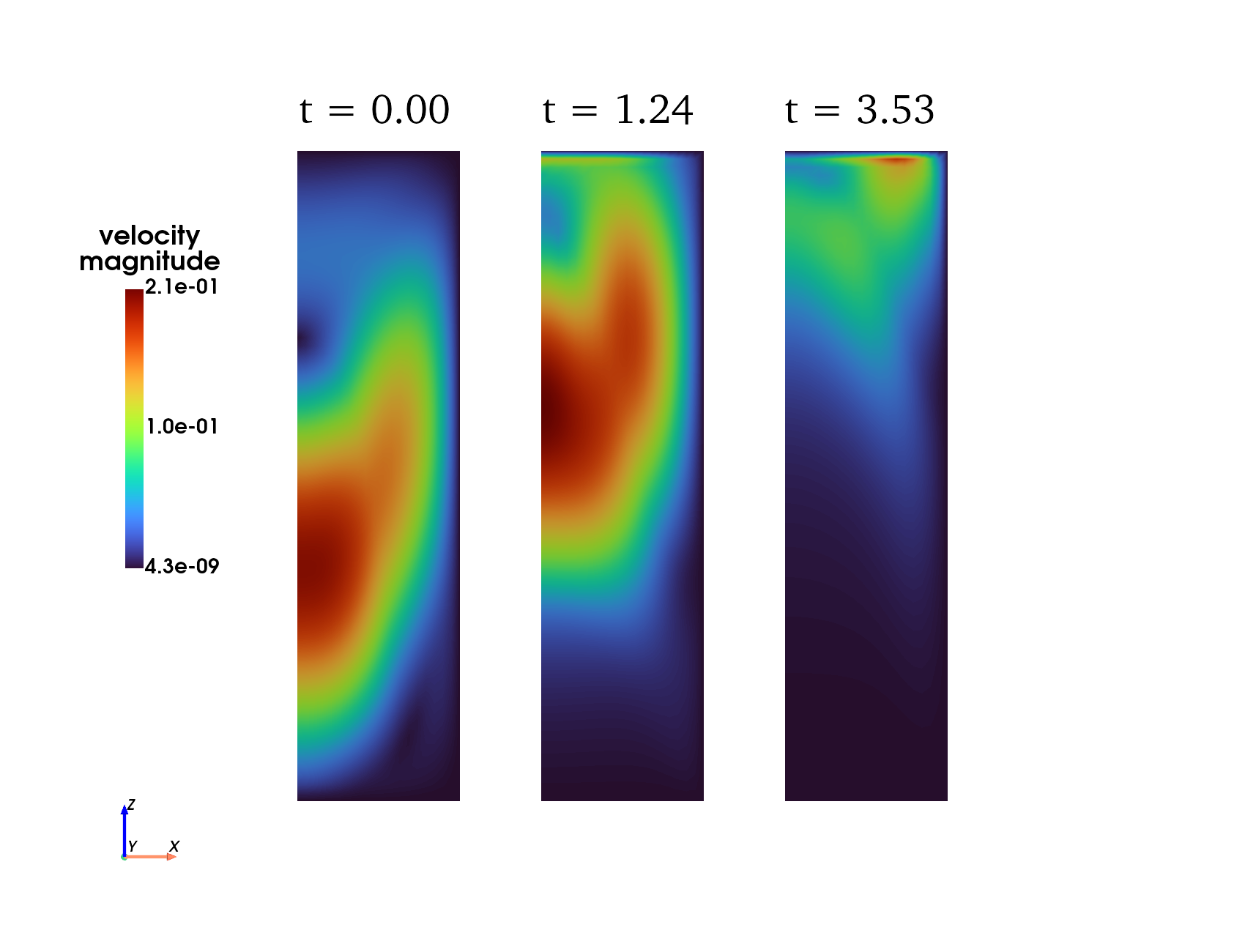}
\caption{Azimuthally averaged perturbation magnitude at various times. From left to right, the snapshots correspond to: the initial time $t=0$, the time when the $L^2$-norm of the perturbation reaches its maximum $t = 1.24$ and a time midway through the decreasing phase $t=3.53$. The simulations were conducted using a mesh size of $h = 0.0625$ and Reynolds number $\RE = 160$. {The left-hand side in each panel corresponds to the centerline of the pipe.}}
\label{fig:Re160_magnitude}
\end{figure}

To give a spatial characterization of the perturbation during its evolution
\Cref{fig:Re160_magnitude} finally shows the azimuthal average of the perturbation magnitude,
for Reynolds number $\RE = 160$, in the $rz-$plane, defined by $(0, 1) \times(0, 4)$, at three specific points in time: the initial time $t=0$, the time when the $L^2-$norm of the perturbation reaches its maximum  $t=1.24$, and a time during the decay phase when the $L^2-$norm is half of its maximum value, at $t = 3.53$. These time points correspond to the dots highlighted in
\Cref{fig:L2norm_perturbation_Rebig}. As expected, the perturbation travels down the pipe in the streamwise direction
until it dissipates. The shape of the perturbation slightly changes as it moves through the pipe before ultimately disappearing.

\section{Conclusion}
\label{sec: conclusion}

In this study we numerically investigated the kinetic energy instability \cite{orr1907stability,ref:SerrinStabilityReOrr}
of pipe flow. We considered finite pipe lengths, with the longest being $L=32$. To address the eigenvalue problem, we imposed either periodic boundary conditions or homogeneous Dirichlet boundary conditions. In the case of periodic boundary conditions, the critical Reynolds number varies periodically with the pipe length, reaching local minima at $\RE_c = 81.58$. Conversely, with homogeneous Dirichlet boundary conditions, the critical Reynolds number decreases as the pipe length increases, suggesting a lower bound smaller than $\RE = 81.62$. Nonetheless, in both scenarios, the critical Reynolds numbers converge to similar values, which are also comparable to those found by Joseph and Carmi \cite{joseph1969stability} ($\RE_c = 81.49$ for periodic perturbations). Unlike other studies \cite{joseph1969stability, falsaperla2019nonlinear}, our approach did not impose assumptions regarding the nature of the perturbations (streamwise or spanwise). The periodic perturbations depend on all the variables, but show a greater dependence on the spanwise ($x$ and $y$) directions. 

We then computed the temporal evolution of the most unstable mode, focusing on the eigenmode with homogeneous Dirichlet boundary conditions. As expected from the theory, when analyzing the perturbed flow at the critical Reynolds number, the $L^2-$norm of the perturbation decreases from the initial value. In contrast, when solving for Reynolds numbers exceeding the critical value, the $L^2-$norm of the perturbation initially increases before eventually decaying. The perturbations slightly change shape over time as they propagate down the pipe until they dissipate. Additionally, the perturbations exhibit a stronger streamwise component during the initial phase of their evolution. 

Our analysis concentrated on Reynolds numbers around $\RE \approx 100$, and as a result, we did not observe
the formation of puffs or slugs \cite{avila2023transition}. Analysis of regimes with higher Reynolds number
is a potentially interesting topic for future research.


\section{{Declaration of generative AI and AI-assisted technologies in the writing process}}

{During the preparation of this work GPT UiO was used in order to improve the readability and language for the parts written by non-native English speakers. After using this tool, the authors reviewed and edited the content as needed and take full responsibility for the content of the published article.}

\section{{Acknowledgments}}
CG acknowledges funding from the European Union's Horizon 2020 research and innovation programme under the Marie Skłodowska-Curie grant agreement 945371.
MK gratefully acknowledges support from the Norwegian Research Council grant 303362 (DataSim).


\bibliographystyle{elsarticle-num} 
\bibliography{article.bib}

\appendix

\section{Method of mappings}\label{Appendix:Map from cylinder to box}
In case of non-polyhedral domain standard finite element (FE) method may converge
suboptimally due to geometric error in approximating the domain. The issue can
be addressed by isoparametric elements which is particularly suitable for general domains.
However, the support in open source finite element libraries might be missing (as is the
case of FEniCS \cite{logg2012automated}). An alternative approach which targets
curved domains with simple-enough parameterization is the method of mappings. 

\begin{figure}
\centering
\includegraphics[width=0.47\textwidth]{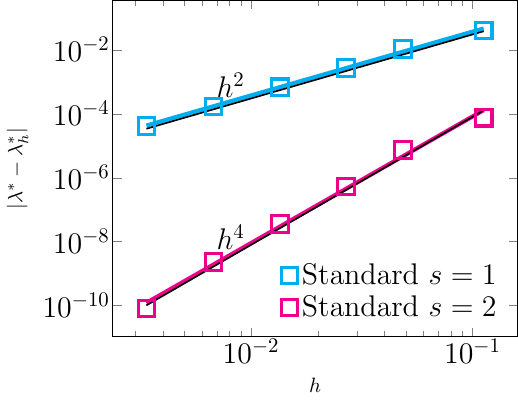}
\hfill
\includegraphics[width=0.47\textwidth]{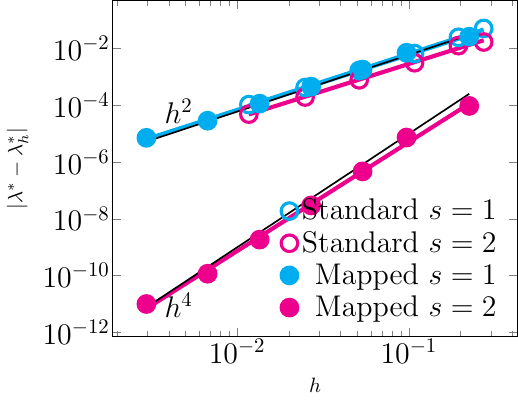}
\caption{Convergence of the finite element method for smallest eigenvalue of the 
problem \eqref{eq:drum} on (left) square domain $\Omega=(-0.5, 0.5)^2$ and (right)
unit disk domain $\Omega$. Continuous Lagrange elements of order $s=1, 2$ are used. 
Standard FE on physical domain leads to optimal $h^{2s}$ convergence only for square domain.
Applying FE on a reference square domain and mapping into physical disk 
domain by \eqref{eq:coordinate_transform} leads to optimal approximation.
}
\label{fig:eig_approx}
\end{figure}

  In order to illustrate the affect of geometry error on the eigenvalue approximation
  we consider the eigenvalue problem for the smallest eigenvalue of the Laplacian with
  homogeneous Dirichlet boundary conditions. That is, we wish to find $\lambda^*=\min \left\{\lambda\right\}$ where
\begin{equation}\label{eq:drum}
    \begin{aligned}
-\Delta u &= \lambda u&\quad\mbox{ in }\Omega,\\
        u &= 0&\quad\mbox{ in }\partial\Omega.\\
\end{aligned}
\end{equation}
Discretizing \eqref{eq:drum} over the finite element space $V^s_h$ of continuous Lagrange elements of
polynomial order $s=1, 2$ leads to the eigenvalue problem: find $\lambda\in\mathbb{R}$, $u\in V^s_h$ such that
\begin{equation}\label{eq:standard_fem}
  \int_{\Omega} \nabla u\cdot \nabla v\,\mathrm{d}\mathbf{x} = \lambda \int_{\Omega} u v\,\mathrm{d}\mathbf{x}\quad\forall v\in V^s_h.
\end{equation}
In \Cref{fig:eig_approx} we observe that the discrete minimal eigenvalue convergences with order $2s$ when
$\Omega=(-0.5, 0.5)^2$. In this case triangulation results in the exact representation of the problem domain. We also note
that the obtained rates are theoretically expected, see \cite[Theorem 10.4]{boffi2010finite}.
On the other hand, with $\Omega$ a unit disk it can be seen 
that the approximation errors decay quadratically for $s=1$ as well as for $s=2$.
The latter case is clearly suboptimal and the slow convergence is due to error in approximating
the curved boundary $\Omega$. 

To address the issue we consider a transformation of coordinates, $\Phi: \hat{\Omega}\rightarrow\Omega$
where $\hat{\Omega}=(-1, 1)^2$. In particular, following \cite{fong2015analytical}, for any point $\mathbf{x}=(x, y)$ in a  
unit disk there exists $\hat{\mathbf{x}}=(\hat{x}, \hat{y})$ such that 
\begin{equation}\label{eq:coordinate_transform}
\mathbf{x}
 = \Phi(\hat{\mathbf{x}})
= \begin{pmatrix}
\hat{x} \sqrt{1- \frac{\hat{y}^2}{2}}\\
\hat{y} \sqrt{1- \frac{\hat{x}^2}{2}}
\end{pmatrix}.
\end{equation}
Using \eqref{eq:coordinate_transform} we can apply finite element method on the reference domain.
That is, with the finite element space $\hat{V}^s_h$ setup over the domain $\hat{\Omega}$ we wish to find
$\lambda\in\mathbb{R}$, $\hat{u}\in \hat{V}^s_h$ satisfying
\begin{equation}\label{eq:mapped_fem}
  \int_{\hat{\Omega}} \hat{J} \left[\hat{\nabla} \hat{u} \cdot \hat{G}^{-1}\right] \cdot
  \left[\hat{\nabla} \hat{v} \cdot \hat{G}^{-1}\right] \,\mathrm{d}\hat{\mathbf{x}}
  = \lambda \int_{\hat{\Omega}} \hat{u} \hat{v} \hat{J}\,\mathrm{d}\hat{\mathbf{x}}\quad\forall \hat{v}\in \hat{V}^s_h.
\end{equation}
Here $\hat{\nabla}=(\partial_{\hat{x}}, \partial_{\hat{y}})$, $\hat{G}=\hat{\nabla}\Phi$ and $\hat{J}=\det \hat{G}$.
As the computational domain $\hat{\Omega}$ is now polygonal it can be discretized without error. However,
comparing \eqref{eq:standard_fem} with \eqref{eq:mapped_fem}, the coordinate
transform yields more involved bilinear forms with integrands which, unlike with computations directly in
the physical domain, are no longer polynomials and thus entail a quadrature error.
In the following we use a quadrature rule exact for
polynomials of degree 10. Applying the coordinate transform to \eqref{eq:drum} we recover convergence rates of
order $2s$ for both linear and quadratic finite elements, see \Cref{fig:eig_approx}.

To apply the mapping method to the Stokes eigenvalue problem \eqref{eq:eigenproblem with lagrange multipliers}
we consider the coordinate transformation
\begin{equation}\label{eq:coordinate_transform_c}
\mathbf{x}
 = \Phi(\hat{\mathbf{x}})
= \begin{pmatrix}
\hat{x} \sqrt{1- \frac{\hat{y}^2}{2}}\\
\hat{y} \sqrt{1- \frac{\hat{x}^2}{2}}\\
\hat{z}
\end{pmatrix}.
\end{equation}
for $\hat{\mathbf{x}}=(\hat{x}, \hat{y}, \hat{z})$ and points $\mathbf{x}$ inside
the cylinder or radius $R=1$.
In \Cref{tab: homo_bc cylinder vs box}, we then report the critical Reynolds numbers for pipe lengths $L=2 \text{ and } 4$ under
homogeneous Dirichlet boundary conditions for various mesh sizes. The results for the periodic boundary conditions are presented
in \Cref{tab: per_bc cylinder vs box}, which also includes critical values for pipe lengths $L=5.87 \text{ and } 11.74$ {($L = k, \, 2k$ with $k$
  defined in \eqref{eq: JC periodicity})}, corresponding to the first two critical lengths reported in \cite{joseph1969stability}. Considering the relative errors, in both cases the method of
mapping appears to lead to more monotone convergence and, in general, the relative error for $h=0.125, 0.0625$ is smaller
than with standard FE.

It is also important to note that the accuracy of the approximated solution using mapped FE 
is sensitive to the tolerance set for the eigenvalue solver residual, as defined in equation
\eqref{eq: eigenproblem residual error}. Optimal results are achieved when this tolerance is
fixed at or below $10^{-8}$.

\begin{table}[h]
  \centering
  \scriptsize
\begin{tabular}{|l|llll||llll|}
\hline
\multicolumn{1}{|l|}{}  & \multicolumn{4}{c||}{Standard}   & \multicolumn{4}{c|}{Mapped} \\ 
\hline
\multicolumn{1}{|l|}{\backslashbox{$L$}{$h$}} & 0.5   & 0.25  & 0.125  & 0.0625 & 0.5   & 0.25  & 0.125  & 0.0625\\
\hline 
\multirow{2}{*}{2} & 102.34 & 100.51 & 98.94  & 98.68 & 100.40 & 99.52 & 98.69 & 98.62 \\
& -- & $1.82E-2$ & $1.58E-2$ & $2.63E-3$ 
& -- & $ 8.84E-3$ & $ 8.41E-3$ & $ 7.09E-4$\\
\hline
\multirow{2}{*}{4} & 85.48 & 83.98 & 83.18 & 82.97 & 81.76 & 83.11 & 82.93 & 82.91\\
& -- & $ 1.78E-2$ & $ 9.61E-3$ & $ 2.53E-3$ 
& -- & $ 1.62E-2$ & $ 2.17E-3$ & $ 2.41E-4$ \\
\hline
\end{tabular}
\caption{Critical Reynolds number and its relative error for different pipe lengths ($L$) and mesh sizes ($h$) computed with standard FE and mapped FE. Homogeneous Dirichlet boundary conditions are prescribed.}
\label{tab: homo_bc cylinder vs box}
\end{table}
%
\begin{table}[h]
  \centering
  \scriptsize
\begin{tabular}{|l|llll||llll|}
\hline
\multicolumn{1}{|l|}{}  & \multicolumn{4}{c||}{Standard}   & \multicolumn{4}{c|}{Mapped} \\ 
\hline
\multicolumn{1}{|l|}{\backslashbox{$L$}{$h$}} & 0.5   & 0.25  & 0.125  & 0.0625 & 0.5   & 0.25  & 0.125  & 0.0625\\
\hline 
\multirow{2}{*}{2} & 83.62 & 83.74 & 83.12 & 82.93 & 82.89 & 82.93 & 82.88 & 82.85 \\
& -- & $1.43E-2$ & $7.45E-3$ & $2.29E-3$ 
& -- & $4.82E-4$ & $ 6.03E-4$ & $ 3.62E-4$ \\
\hline
\multirow{2}{*}{4} & 83.73 & 83.75 & 82.84 & 82.63 & 79.87 & 82.89 & 82.60 & 82.57 \\
& -- &  $2.38E-4$ & $ 1.09E-2$ & $ 2.54E-3$ 
& -- &  $3.64E-2$ & $ 3.5E-3$ & $ 3.63E-4$ \\
\hline
\multirow{2}{*}{5.87} & 83.52 & 82.58 & 81.78  & 81.58 & 79.73 & 81.71 & 81.53 & 81.52 \\
& -- &  $ 1.13E-2$ & $ 9.78E-3$ & $ 2.45E-3$ 
& -- &  $2.42E-2$ & $ 2.20E-3$ & $ 1.22E-4$\\
\hline
\multirow{2}{*}{11.74} & 82.93 & 82.66 & 81.78  & 81.58 & 79.25 & 81.73 & 81.54 & 81.52 \\
& -- &  $3.26E-3$ & $ 1.07E-2$ & $ 2.45E-3$
& -- &  $ 3.03E-2$ & $ 2.33E-3$ & $ 2.45E-4$\\
\hline
\end{tabular}
\caption{Critical Reynolds number and its relative error for different pipe lengths ($L$) and mesh sizes ($h$) computed with standard FE and mapped FE. Periodic boundary conditions are prescribed on inlet/outlet.}
\label{tab: per_bc cylinder vs box}
\end{table}

\end{document}